\documentclass[aps,preprint,showpacs,groupedaddress,floatfix]{revtex4}
\usepackage{graphicx}
\usepackage{dcolumn}
\usepackage{bm}

\begin{document}

\title{Quasiparticle random phase approximation based on the relativistic
Hartree-Bogoliubov model}
\author{N. Paar}
\affiliation{Physik-Department der Technischen Universit\"at M\"unchen, D-85748 Garching,
Germany}
\author{T. Nik\v si\' c}
\author{D. Vretenar}
\affiliation{Physics Department, Faculty of Science, University of Zagreb, Croatia, and \\
Physik-Department der Technischen Universit\"at M\"unchen, D-85748 Garching,
Germany}
\author{P. Ring}
\affiliation{Physik-Department der Technischen Universit\"at M\"unchen, D-85748 Garching,
Germany}
\date{\today}

\begin{abstract}
The relativistic quasiparticle random phase approximation (RQRPA) is
formulated in the canonical single-nucleon basis of the relativistic
Hartree-Bogoliubov (RHB) model. For the interaction in the particle-hole
channel effective Lagrangians with nonlinear meson self-interactions are
used, and pairing correlations are described by the pairing part of the
finite range Gogny interaction. The RQRPA configuration space includes the
Dirac sea of negative energy states. Both in the particle-hole and
particle-particle channels, the same interactions are used in the RHB
calculation of the ground state and in the matrix equations of the RQRPA.
The RHB+RQRPA approach is tested in the example of multipole excitations of
neutron rich oxygen isotopes. The RQRPA is applied in the analysis of the
evolution of the low-lying isovector dipole strength in Sn isotopes and N=82
isotones.
\end{abstract}

\maketitle

\bigskip \bigskip

\section{\label{secI}Introduction}


The multipole response of unstable nuclei far from the line of $\beta$%
-stability presents a very active field of research, both experimental and
theoretical. These nuclei are characterized by unique structure properties:
the weak binding of the outermost nucleons and the effects of the coupling
between bound states and the particle continuum. On the neutron rich side,
in particular, the modification of the effective nuclear potential leads to
the formation of nuclei with very diffuse neutron densities, to the
occurrence of the neutron skin and halo structures. These phenomena will
also affect the multipole response of unstable nuclei, in particular the
electric dipole and quadrupole excitations, and new modes of excitations
might arise in nuclei near the drip line.

A quantitative description of ground-states and properties of excited states
in nuclei characterized by the closeness of the Fermi surface to the
particle continuum, necessitates a unified description of mean-field and
pairing correlations, as for example in the framework of the
Hartree-Fock-Bogoliubov (HFB) theory. In order to describe transitions to
low-lying excited states in weakly bound nuclei, in particular, the
two-quasiparticle configuration space must include states with both nucleons
in the discrete bound levels, states with one nucleon in a bound level and
one nucleon in the continuum, and also states with both nucleons in the
continuum. This cannot be accomplished in the framework of the BCS
approximation, since the BCS scheme does not provide a correct description
of the scattering of nucleonic pairs from bound states to the positive
energy particle continuum. Collective low-lying excited states in weakly
bound nuclei are best described by the quasiparticle random phase
approximation (QRPA) based on the HFB framework. The HFB based QRPA has been
investigated in a number of recent theoretical studies. In Ref.~\cite{Eng.99}
a fully self-consistent QRPA has been formulated in the HFB canonical
single-particle basis. The Hartree-Fock-Bogoliubov formalism in coordinate
state representation has also been used as a basis for the continuum linear
response theory~\cite{Mat.01,Mat.02}. In Ref.~\cite{Kha.02} the HFB energy
functional has been used to derive the continuum QRPA response function in
coordinate space. HFB based continuum QRPA calculations have been performed
for the low-lying excited states and giant resonances, as well as for the $%
\beta$ decay rates in neutron rich nuclei.

In this work we formulate the relativistic QRPA in the canonical
single-nucleon basis of the relativistic Hartree-Bogoliubov (RHB) model. The
RHB model is based on the relativistic mean-field theory and on the
Hartree-Fock-Bogoliubov framework. It has been very successfully applied in
the description of a variety of nuclear structure phenomena, not only in
nuclei along the valley of $\beta$-stability, but also in exotic nuclei with
extreme isospin values and close to the particle drip lines. Another
relativistic model, the relativistic random phase approximation (RRPA), has
been recently employed in quantitative analyses of collective excitations in
nuclei. Two points are essential for the successful application of the RRPA
in the description of dynamical properties of finite nuclei: (i) the use of
effective Lagrangians with nonlinear self-interaction terms, and (ii) the
fully consistent treatment of the Dirac sea of negative energy states.

The RRPA with nonlinear meson interaction terms, and with a configuration
space that includes the Dirac sea of negative-energy state, has been very
successfully employed in studies of nuclear compressional modes~\cite%
{Vre.00,Pie.01,MGW.01}, of multipole giant resonances and of low-lying
collective states in spherical nuclei~\cite{Mawa.01}, of the evolution of
the low-lying isovector dipole response in nuclei with a large neutron
excess~\cite{Vrepyg1.01,Vrepyg2.01}, and of toroidal dipole resonances~\cite%
{Vre.02}.

In Sec.~\ref{secII} we present the formalism and formulate the matrix
equations of the relativistic QRPA in the canonical basis of the
relativistic Hartree-Bogoliubov (RHB) framework for spherical even-even
nuclei. In Sec.~\ref{secIII} the RHB+RQRPA approach is tested in the example
of the isoscalar monopole, isovector dipole and isoscalar quadrupole
response of $^{22}$O, and the results are compared with recent
non-relativistic QRPA calculations of the multipole response of neutron rich
oxygen isotopes. In Sec.~\ref{secIV} the RQRPA framework is applied in the
analysis of the evolution of the low-lying isovector dipole strength in Sn
isotopes and N=82 isotones. The results are compared with recent
experimental data. Section~\ref{secV} contains the summary and the
conclusions.

\section{\label{secII}The relativistic quasiparticle random phase
approximation}


In this section the matrix equations of the relativistic quasiparticle
random phase approximation (RQRPA) are formulated in the canonical basis of
the relativistic Hartree-Bogoliubov (RHB) framework for spherical even-even
nuclei.

\subsection{The relativistic mean-field Lagrangian and the equations of
motion}

The nuclear matter equation of state and detailed properties of finite
nuclei have been very successfully described by relativistic mean-field
(RMF) models~\cite{Ser.86,Rei.89,Rin.96}. In this framework the nucleus is
described as a system of Dirac nucleons that interact in a relativistic
covariant manner by meson exchange. In particular, the isoscalar scalar $%
\sigma$-meson, the isoscalar vector $\omega$-meson, and the isovector vector
$\rho$-meson build the minimal set of meson fields that is necessary for a
quantitative description of bulk and single-particle nuclear properties. The
model is defined by the Lagrangian density
\begin{equation}
\mathcal{L}=\mathcal{L}_{N}+\mathcal{L}_{m}+\mathcal{L}_{int}.
\label{lagrangian}
\end{equation}%
$\mathcal{L}_{N}$ denotes the Lagrangian of the free nucleon
\begin{equation}
\mathcal{L}_{N}=\bar{\psi}\left( i\gamma ^{\mu }\partial _{\mu }-m\right)
\psi ,
\end{equation}
where $m$ is the bare nucleon mass and $\psi$ denotes the Dirac spinor. $%
\mathcal{L}_{m}$ is the Lagrangian of the free meson fields and the
electromagnetic field
\begin{eqnarray}
\mathcal{L}_{m} &=&\frac{1}{2}\partial_{\mu }\sigma \partial^{\mu }\sigma -
\frac{1}{2}m_{\sigma }^{2}\sigma ^{2}-\frac{1}{4}\Omega_{\mu \nu }\Omega
^{\mu \nu }+\frac{1}{2}m_{\omega }^{2}\omega _{\mu }\omega ^{\mu }  \nonumber
\\
&&-\frac{1}{4}\vec{R}_{\mu \nu }\vec{R}^{\mu \nu }+\frac{1}{2}m_{\rho }^{2}%
\vec{\rho}_{\mu }\vec{\rho}^{\mu }-\frac{1}{4}F_{\mu \nu }F^{\mu \nu } ,
\end{eqnarray}
with the corresponding masses $m_{\sigma }$, $m_{\omega }$, $m_{\rho }$, and
$\Omega_{\mu\nu}$, $\vec{R}_{\mu\nu}$, $F_{\mu\nu}$ are field tensors
\begin{equation}
\begin{array}{rcl}
\Omega _{\mu \nu } & = & \partial _{\mu }\omega _{\nu }-\partial _{\nu
}\omega _{\mu } \\
\vec{R}_{\mu \nu } & = & \partial _{\mu }\vec{\rho}_{\nu }-\partial _{\nu }%
\vec{\rho}_{\mu } \\
F_{\mu \nu } & = & \partial _{\mu }{A}_{\nu }-\partial _{\nu }{A}_{\mu }.%
\end{array}%
\end{equation}%
The model Lagrangian density contains also the interaction terms
\begin{equation}
\mathcal{L}_{int}=-\bar{\psi}\Gamma _{\sigma }\sigma \psi -\bar{\psi}\Gamma
_{\omega }^{\mu}\omega_{\mu} \psi -\bar{\psi} \vec{\Gamma}_{\rho }^{\mu}\vec{%
\rho}_{\mu}\psi -\bar{\psi}\Gamma _{e}^{\mu}A_{\mu}\psi .
\end{equation}%
The vertices read
\begin{equation}
\Gamma_{\sigma }=g_{\sigma },\;\;\;\Gamma_{\omega }^{\mu }=g_{\omega }\gamma
^{\mu },\;\;\;\vec{\Gamma}_{\rho }^{\mu }=g_{\rho }\vec{\tau}\gamma ^{\mu
},\;\;\;\;\Gamma_{e}^{m}=e\frac{1-\tau _{3}}{2}\gamma ^{\mu },
\end{equation}%
with the coupling constants $g_{\sigma }$, $g_{\omega }$, $g_{\rho }$ and $e$%
. This simple linear model, however, does not provide a quantitative
description of complex nuclear systems. An effective density dependence has
been introduced~\cite{Bog.77} by replacing the quadratic $\sigma$-potential $%
\displaystyle \frac{1}{2}m_{\sigma }^{2}\sigma ^{2}$ with a quartic
potential $\displaystyle U(\sigma )=\frac{1}{2}m_{\sigma }^{2}\sigma ^{2}+%
\frac{g_{2}}{3}\sigma ^{3} +\frac{g_{3}}{4}\sigma ^{4}$. This potential
includes the nonlinear $\sigma$ self-interactions with two additional
parameters $g_2$ and $g_3$. It has been shown that one can describe the
properties of nuclear matter and finite nuclei with high accuracy using
density dependent coupling constants $g_{m}(\rho )$, instead of nonlinear $%
\sigma$ self-interaction~\cite{TDP.02}.

From the model Lagrangian density the classical variation principle leads to
the equations of motion. The time-dependent Dirac equation for the nucleon
reads%
\begin{equation}
\left[ \gamma ^{\mu }(i\mathbf{\partial }_{\mu }+V_{\mu })+m+S\right] \psi
=0.  \label{Dirac0}
\end{equation}%
If one neglects retardation effects for the meson fields, a self-consistent
solution is obtained when the time-dependent mean-field potentials
\begin{eqnarray}
S(\mathbf{r},t) &=&g_{\sigma }\sigma (\mathbf{r},t)~,  \nonumber \\
V_{\mu }(\mathbf{r},t) &=&g_{\omega }\omega _{\mu }(\mathbf{r},t)+g_{\rho }%
\vec{\tau}\vec{\rho}_{\mu }(\mathbf{r},t)+eA_{\mu }(\mathbf{r},t)\frac{%
(1-\tau _{3})}{2}~,  \label{Potentials}
\end{eqnarray}%
are calculated at each step in time from the solution of the stationary
Klein-Gordon equations
\begin{equation}
-\Delta \phi _{m}+U^{\prime }(\phi _{m})=\pm \left\langle \bar{\psi}\Gamma
_{m}\psi \right\rangle ,  \label{KG0}
\end{equation}%
where the upper sign holds for vector fields and the lower sign for the
scalar field. The index $m$ denotes mesons and the photon, i.e. $\phi
_{m}=\sigma ,\omega ^{\mu },\vec{\rho}^{\mu },A^{\mu }$. This approximation
is justified by the large meson masses. The corresponding meson exchange
forces are of short range and therefore retardation effects can be neglected.

In practical applications to nuclear matter and finite nuclei, the
relativistic models are used in the \textit{no-sea} approximation: the Dirac
sea of states with negative energies does not contribute to the densities
and currents. For a nucleus with A nucleons
\begin{equation}
\left\langle\bar{\psi}\Gamma_{m}\psi\right\rangle = \sum\limits_{i=1}^{A}%
\bar{\psi}_{i}^{{}}(\mathbf{r},t)\Gamma_m \psi _{i}^{{}}(\mathbf{r},t)\; ,
\label{Densities}
\end{equation}
where the summation is performed only over the occupied orbits in the Fermi
sea of positive energy states. The set of coupled equations (\ref{Dirac0})
and (\ref{KG0}) define the relativistic mean field (RMF) model. In the
stationary case they reduce to a nonlinear eigenvalue problem, and in the
time-dependent case they describe the nonlinear propagation of the Dirac
spinors in time \cite{VBR.95}.

The mean-field approximation represents the lowest order of the quantum
field theory: the meson field operators are replaced by their expectation
values. The A nucleons, described by a Slater determinant $|\Phi \rangle $
of single-particle spinors, move independently in the classical meson
fields. The couplings of the meson fields to the nucleon are adjusted to
reproduce the properties of nuclear matter and finite nuclei. The $\sigma$%
-meson approximates a large attractive scalar field, that is produced by
very complicated microscopic processes, such as uncorrelated and correlated
pion-exchange. The $\omega$-meson describes the short range repulsion
between the nucleons, and the $\rho$-meson carries the isospin quantum
number. The latter is required by the large empirical asymmetry potential in
finite nuclear systems. The basic ingredient of the microscopic nuclear
force is the pion. In relativistic mean-field models it does not contribute
on the Hartree level because of parity conservation. The pion field has been
included in the relativistic Hartree-Fock model. However, the resulting
equations of motion are rather complicated and this model has been rarely used. Many
effects that go beyond the mean-field level are apparently neglected in the
RMF model. Among them are the Fock terms, the vacuum polarization effects
and the short range Brueckner-type correlations. The experimental data to
which the meson-nucleon couplings are adjusted, however, contain all these
effects and much more. It follows that these effects are not completely
neglected. On the contrary, they are taken into account in an effective way.
The concept behind the RMF model is therefore equivalent to that of the
density functional theory, which is widely used in solid state physics,
molecular physics, chemistry and also in non-relativistic nuclear physics.
The RMF model represent the covariant form of this method.


\subsection{Covariant density functional theory}

The equations of motion of the relativistic mean field model can also be
derived starting from a density functional. From the energy-momentum tensor
one writes the total energy of the nuclear system
\begin{eqnarray}
E_{RMF}[\psi, \bar{\psi}, \sigma, \omega^{\mu}, \vec{\rho}^{\mu}, A^{\mu}]
&=& \sum_{i=1}^{A}\int \psi_i^{+}({\mbox{\boldmath $\alpha$}} {%
\mbox{\boldmath $p$}}+\beta m)\psi_i  \nonumber \\
&+& \int \left[ \frac{1}{2}(\nabla \sigma )^2 + U(\sigma ) \right]d^3r -
\frac{1}{2} \int \left[ (\nabla \omega )^2 + m_{\omega}^2\omega^2 ) \right]%
d^3r  \nonumber \\
&-&\frac{1}{2} \int \left[ (\nabla \rho )^2 + m_{\rho}^2\rho^2 ) \right]d^3r
-\frac{1}{2} \int (\nabla A )^2 d^3r  \nonumber \\
&+& \int \left[ g_{\sigma}\rho_s \sigma + g_{\omega}j_{\mu}\omega^{\mu} +
g_{\rho}\vec{j}_{\mu}\vec{\rho}^{\mu}+e j_{c\mu}A^{\mu} \right] d^3r.
\label{E-RMF0}
\end{eqnarray}%
By using the definition of the relativistic single nucleon density matrix
\begin{equation}
\hat{\rho}(\mathbf{r,r}^{\prime },t)=\sum\limits_{i=1}^{A}|\psi _{i}(\mathbf{%
r,}t)\rangle \langle \psi _{i}(\mathbf{r}^{\prime },t)| \;,
\label{density-HF}
\end{equation}%
the total energy can be written as a functional of the density matrix $\hat{%
\rho}$ and the meson fields
\begin{eqnarray}
E_{RMF}[\hat{\rho},\phi_m ] &=&\text{Tr}\left[ ({\mbox{\boldmath $\alpha$}}{%
\mbox{\boldmath $p$}}+\beta m) \hat{\rho}\right]  \nonumber \\
&&\pm\int \left[ \frac{1}{2}(\mathbf{\nabla}\phi_m)^{2} +U(\phi_m) \right]%
d^{3}r  \nonumber \\
&&+\text{Tr}\left[ (\Gamma _{m}\phi _{m})\hat{\rho}\right] .  \label{E-RMF}
\end{eqnarray}%
The trace operation involves a sum over the Dirac indices and an integral in
coordinate space. The index m is used as generic notation for all mesons and
the photon. From the classical time-dependent variational principle%
\begin{equation}
\delta \int_{t_{1}}^{t_{2}}dt\left\{ \left\langle \Phi |i\partial _{t}|\Phi
\right\rangle -E\left[ \hat{\rho},\phi_m \right] \right\} =0
\label{actionprinciple}
\end{equation}%
the equations of motion (\ref{Dirac0}) and (\ref{KG0}) are obtained. The
equation of motion for the density matrix reads
\begin{eqnarray}
i\partial _{t}\hat{\rho} &=&\left[ \hat{h}(\hat{\rho},\phi_m ), \hat{\rho}%
\right] \;.  \label{hr-rh}
\end{eqnarray}%
The single particle Hamiltonian $\hat{h}$ is the functional derivative of
the energy with respect to the single particle density matrix $\hat{\rho}$
\begin{equation}
\hat{h}=\frac{\delta E}{\delta \hat{\rho}} .
\end{equation}

\subsection{Pairing correlations and the Relativistic Hartree-Bogoliubov
theory}

The inclusion of pairing correlations is essential for a quantitative
description open shell nuclei. In Ref. \cite{KR.91} a fully microscopic
derivation of the relativistic Hartree-Bogoliubov theory has been developed.
Using the Gorkov factorization technique, it has been shown that the pairing
interaction results from the one-meson exchange ($\sigma $-, $\omega $- and $%
\rho $-mesons). In practice, however, it turns out that the pairing
correlations calculated in this way, with coupling constants taken from the
standard parameter sets of the RMF model, are too strong. The repulsion
produced by the exchange of vector mesons at short distances results in a
pairing gap at the Fermi surface that is by a factor three too large.
However, as has been argued in many applications of the
Hartree-Fock-Bogoliubov theory, there is no real reason to use the same
effective forces in both the particle-hole and particle-particle channels.

Pairing correlations can be easily included in the framework of density
functional theory, by using a generalized Slater determinant $|\Phi \rangle $
of the Hartree-Bogoliubov type. The ground state of a nucleus $|\Phi \rangle
$ is represented as the vacuum with respect to independent quasiparticle
operators
\begin{equation}
\alpha _{k}^{+}=\sum\limits_{l}U_{lk}^{{}}c_{l}^{+}+V_{lk}^{{}}c_{l}^{{}},
\end{equation}%
where $U_{lk}$, $V_{lk}$ are the Hartree-Bogoliubov coefficients. They
determine the Hermitian single particle density matrix%
\begin{equation}
\hat{\rho}=V^{\ast }V^{T},\;  \label{rho0}
\end{equation}%
and the antisymmetric pairing tensor
\begin{equation}
\hat{\kappa}=V^{\ast }U^{T}.  \label{kappa0}
\end{equation}%
The energy functional depends not only on the density matrix $\hat{\rho}$
and the meson fields $\phi _{m}$, but in addition also on the pairing
tensor. It has the form%
\begin{equation}
E[\hat{\rho},\hat{\kappa},\phi _{m}]=E_{RMF}[\hat{\rho},\phi _{m}]+E_{pair}[%
\hat{\kappa}],  \label{ERHB}
\end{equation}%
where $E_{RMF}[\hat{\rho},\phi ]$ is the $RMF$-functional defined in Eq. (%
\ref{E-RMF}). The pairing energy $E_{pair}[\hat{\kappa}]$ is given by%
\begin{equation}
E_{pair}[\hat{\kappa}]=\frac{1}{4}\text{Tr}\left[ \hat{\kappa}^{\ast }V^{pp}%
\hat{\kappa}\right] .
\end{equation}%
$V^{pp}$ is a general two-body pairing interaction. Finally, the total
energy can be written as a functional of the generalized density matrix \cite%
{Va.61}
\begin{equation}
\mathcal{R}=\left(
\begin{array}{cc}
\rho  & \kappa  \\
-\kappa ^{\ast } & 1-\rho ^{\ast }%
\end{array}%
\right) \;,  \label{R}
\end{equation}%
which obeys the equation of motion%
\begin{equation}
i\partial _{t}\mathcal{R}=\left[ \mathcal{H}(\mathcal{R}),\mathcal{R}\right]
.  \label{TDRHB}
\end{equation}%
The generalized Hamiltonian $\mathcal{H}$ is a functional derivative of the
energy with respect to the generalized density
\begin{equation}
\mathcal{H}~=~\frac{\delta E}{\delta \mathcal{R}}~=~\left(
\begin{array}{cc}
\hat{h}_{D}-m-\lambda  & \hat{\Delta} \\
-\hat{\Delta}^{\ast } & -\hat{h}_{D}+m+\lambda
\end{array}%
\right) \;.  \label{HFB-hamiltonian}
\end{equation}%
It contains two average potentials: the self-consistent mean field $\hat{h}%
_{D}$, which encloses all the long range particle-hole (\textit{ph})
correlations, and the pairing field $\hat{\Delta}$, which includes the
particle-particle (\textit{pp}) correlations. The single particle potential $%
\hat{h}_{D}$ results from the variation of the energy functional with
respect to the Hermitian density matrix $\hat{\rho}$
\begin{equation}
\hat{h}_{D}=\frac{\delta E}{\delta \hat{\rho}},\;
\end{equation}%
and the pairing field is obtained from the variation of the energy
functional with respect to the pairing tensor
\begin{equation}
\hat{\Delta}~=~\frac{\delta E}{\delta \hat{\kappa}}.
\end{equation}%
The pairing field is an integral operator with the kernel
\begin{equation}
\Delta _{ab}(\mathbf{r},\mathbf{r}^{\prime })={\frac{1}{2}}%
\sum\limits_{c,d}V_{abcd}^{pp}(\mathbf{r},\mathbf{r}^{\prime })\mathbf{%
\kappa }_{cd}(\mathbf{r},\mathbf{r}^{\prime }),  \label{equ.2.5}
\end{equation}%
where $a,b,c,d$ denote quantum numbers that specify the Dirac indices of the
spinors, and $V_{abcd}^{pp}(\mathbf{r},\mathbf{r}^{\prime })$ are the matrix
elements of a general two-body pairing interaction.

The stationary limit of Eq. (\ref{TDRHB}) describes the ground state of an
open-shell nucleus~\cite{Gon.96,Lal1.98}. It is determined by the solutions
of the Hartree-Bogoliubov equations%
\begin{equation}
\left(
\begin{array}{cc}
\hat{h}_{D}-m-\lambda & \hat{\Delta} \\
-\hat{\Delta}^{\ast } & -\hat{h}_{D}+m+\lambda%
\end{array}%
\right) \left(
\begin{array}{c}
U_{k}(\mathbf{r}) \\
V_{k}(\mathbf{r})%
\end{array}%
\right) =E_{k}\left(
\begin{array}{c}
U_{k}(\mathbf{r}) \\
V_{k}(\mathbf{r})%
\end{array}%
\right) \;.  \label{eqhb}
\end{equation}%
The chemical potential $\lambda $ is determined by the particle number
subsidiary condition in order that the expectation value of the particle
number operator in the ground state equals the number of nucleons. The
column vectors denote the quasiparticle wave functions, and $E_k$ are the
quasiparticle energies. The dimension of the RHB matrix equation is two
times the dimension of the corresponding Dirac equation. For each
eigenvector $(U_k ,V_k )$ with positive quasiparticle energy $E_k > 0$,
there exists an eigenvector $(V_k^*,U_k^*)$ with quasiparticle energy $-E_k$%
. Since the baryon quasiparticle operators satisfy fermion commutation
relations, the levels $E_k$ and $-E_k$ cannot be occupied simultaneously.
For the solution that corresponds to a ground state of a nucleus with even
particle number, one usually chooses the eigenvectors with positive
eigenvalues $E_k$.

The RHB equations are solved self-consistently, with potentials determined
in the mean-field approximation from solutions of static Klein-Gordon
equations
\begin{eqnarray}
\left[ -\Delta +m_{\sigma }^{2}\right] \,\sigma (\mathbf{r}) &=&-g_{\sigma
}\,\rho _{s}(\mathbf{r})-g_{2}\,\sigma ^{2}(\mathbf{r})-g_{3}\,\sigma ^{3}(%
\mathbf{r})  \label{messig} \\
\left[ -\Delta +m_{\omega }^{2}\right] \,\omega ^{0}(\mathbf{r})
&=&g_{\omega }\,\rho _{v}(\mathbf{r})  \label{mesome} \\
\left[ -\Delta +m_{\rho }^{2}\right] \,\rho ^{0}(\mathbf{r}) &=&g_{\rho
}\,\rho _{3}(\mathbf{r})  \label{mesrho} \\
-\Delta \,A^{0}(\mathbf{r}) &=&e\,\rho _{p}(\mathbf{r})  \label{photon}
\end{eqnarray}%
for the $\sigma $-meson, the $\omega $-meson, the $\vec{\rho}$-meson and the
photon field, respectively. Because of charge conservation, only the 3-rd
component of the isovector $\rho $-meson contributes. In the ground-state
solution for an even-even nucleus there are no currents (time reversal
invariance) and the spatial components ${\mbox{\boldmath $\omega $}}$, ${%
\mbox{\boldmath $\rho$}}_{3}$, ${\mbox{\boldmath $A$}}$ of the vector fields
vanish. In nuclei with an odd number of protons or neutrons time reversal
symmetry is broken, and the resulting spatial components of the meson fields
play an essential role in the description of magnetic moments and of moments
of inertia in rotating nuclei. The equation for the isoscalar scalar $\sigma
$-meson field contains nonlinear terms. The inclusion of nonlinear meson
self-interaction terms in meson-exchange RMF models is absolutely necessary
for a quantitative description of ground-state properties of spherical and
deformed nuclei~\cite{Rin.96}. The source terms in equations (\ref{messig})
to (\ref{photon}) are sums of bilinear products of baryon amplitudes
\begin{eqnarray}
\rho _{s}(\mathbf{r}) &=&\sum\limits_{k>0}V_{k}^{\dagger }(\mathbf{r})\gamma
^{0}V_{k}(\mathbf{r})  \label{densities} \\
\rho _{v}(\mathbf{r}) &=&\sum\limits_{k>0}V_{k}^{\dagger }(\mathbf{r})V_{k}(%
\mathbf{r}) \\
\rho _{3}(\mathbf{r}) &=&\sum\limits_{k>0}V_{k}^{\dagger }(\mathbf{r})\tau
_{3}V_{k}(\mathbf{r})_{{}} \\
\rho _{\mathrm{em}}(\mathbf{r}) &=&\sum\limits_{k>0}V_{k}^{\dagger }(\mathbf{%
r}){\frac{{1-\tau _{3}}}{2}}V_{k}(\mathbf{r})\;,
\end{eqnarray}%
where $\sum_{k>0}$ is a shorthand notation for the \textit{no-sea}
approximation. The self-consistent solution of the Dirac-Hartree-Bogoliubov
integro-differential equations and Klein-Gordon equations for the meson
fields determines the ground state of a nucleus. In the present
implementation of the RHB model the coupled system of equations is solved by
expanding the nucleon spinors $U_{k}(\mathbf{r})$ and $V_{k}(\mathbf{r})$,
and the meson fields in the spherical harmonic oscillator basis~\cite{GRT.90}.

\bigskip


\subsection{The Relativistic Quasi-Particle Random Phase Approximation}

In this section we will derive the RQRPA from the time-dependent RHB model
in the limit of small amplitude oscillations. The 
generalized density matrix $\mathcal{R}$ and the fields
$\phi_m=\sigma ,\omega ^{\mu },\vec{\rho}^{\mu },A^{\mu }$ 
have been considered as independent variables related only by the
equations of motion. One can use the Klein-Gordon equations to eliminate the
meson degrees of freedom, but this is only possible in the small amplitude
limit. The time-dependent meson field can be written as
\begin{equation}
\phi_m~=~\phi^{(0)}_m + \delta\phi_m,
\end{equation}
where $\phi^{(0)}_m$ is the meson field that corresponds to the stationary
ground state, and $\delta\phi_m$ is a small variation of the meson field
around the stationary state solution. In the linear approximation the
corresponding Klein-Gordon equation reads
\begin{equation}
\left[-\Delta + U^{\prime\prime}(\phi_m^{(0)})\right]\delta\phi_{m}(\mathbf{r%
}) =\pm g_m\delta\rho_m(\mathbf{r}),
\end{equation}
where $\delta\rho_m(\mathbf{r})$ are the various densities and currents (see
Eq. (\ref{Densities})). If there are no nonlinear meson self-interaction
terms, $U^{\prime\prime}(\phi_m^{(0)})=m^2_m$. The propagator $G_m(\mathbf{r}%
,\mathbf{r}^{\prime})$ can be obtained analytically and it has the Yukawa
form. In the case of nonlinear meson self-interaction terms $%
U^{\prime\prime}(\phi_m^{(0)})$ depends on the field $\phi_m^{(0)}$, and an
analytical solution is no longer possible. The propagator $G_m(\mathbf{r},%
\mathbf{r}^{\prime})$ has to be calculated numerically (for details see Ref.
\cite{MGT.97}). In both cases we find a linear relation between $%
\delta\phi_m $ and $\delta{\rho}_m$
\begin{equation}
\delta\phi_m(\mathbf{r})=\pm g_m\int d^3r^{\prime}G_m(\mathbf{r},\mathbf{r}%
^{\prime})\delta\rho_m(\mathbf{r}^{\prime}).
\end{equation}

The generalized Hamiltonian $\mathcal{H}$ can now be expressed as a
functional of the generalized density $\mathcal{R}$ only. In the linear
approximation the generalized density matrix is expanded
\begin{equation}
\mathcal{R}=\mathcal{R}_{0}+\delta \mathcal{R}(t),  \label{lindens}
\end{equation}%
where $\mathcal{R}_{0}$ is the stationary ground-state generalized density.
Since $\mathcal{R}(t)$ is a projector at all times, in linear order
\begin{equation}
\mathcal{R}_{0}\delta \mathcal{R}+\delta \mathcal{R}\mathcal{R}_{0}=\delta
\mathcal{R}.  \label{projector}
\end{equation}%
In the quasiparticle basis the matrices $\mathcal{R}_{0}$ and $\mathcal{H}%
_{0}=\mathcal{H}(\mathcal{R}_{0})$ are diagonal
\begin{equation}
\mathcal{R}_{0}=\left(
\begin{array}{cc}
0 & 0 \\
0 & 1%
\end{array}%
\right) ~\ \;\;\;\;\text{and \ ~~~}\mathcal{H}_{0}=\left(
\begin{array}{cc}
E_{n} & 0 \\
0 & -E_{n}%
\end{array}%
\right) .
\end{equation}%
From Eq. (\ref{projector}) it follows that the matrix $\delta \mathcal{R}$
has the form
\begin{equation}
\delta \mathcal{R}=\left(
\begin{array}{cc}
0 & \delta R \\
-\delta R^{\ast } & 0%
\end{array}%
\right) .  \label{deltaR}
\end{equation}%
The linearized equation of motion (\ref{TDRHB}) reduces to
\begin{equation}
i\partial _{t}\mathcal{R}=\left[ \mathcal{H}_{0},\delta \mathcal{R}\right] +%
\left[ \frac{\delta \mathcal{H}}{\delta \mathcal{R}}\delta \mathcal{R},%
\mathcal{R}_{0}\right] .  \label{LTDRHB}
\end{equation}%
Assuming an oscillatory solution
\begin{equation}
\delta \mathcal{R}(t)=\sum_{\nu }\delta \mathcal{R}^{(\nu )}e^{i\omega _{\nu
}t}+h.c.,
\end{equation}%
the RQRPA equation is obtained
\begin{equation}
\left(
\begin{array}{cc}
A & B \\
-B^{\ast } & -A^{\ast }%
\end{array}%
\right) \left(
\begin{array}{c}
X^{\nu } \\
Y^{\nu }%
\end{array}%
\right) =\omega _{\nu }\left(
\begin{array}{c}
X^{\nu } \\
Y^{\nu }%
\end{array}%
\right) .  \label{RQRPA}
\end{equation}%
For $k<k^{\prime }$, $l<l^{\prime }$ the RQRPA matrix elements read%
\begin{equation}
A_{kk^{\prime }ll^{\prime }}=(E_{k^{{}}}+E_{k^{\prime }})\delta
_{k^{{}}l^{{}}}\delta _{k^{\prime }l^{\prime }}+\frac{\delta ^{2}E}{\delta
R_{kk^{\prime }}^{\ast }\delta R_{ll^{\prime }}^{{}}}\text{ \ \ \ \ \ \ \
and \ \ \ \ \ \ }B_{kk^{\prime }ll^{\prime }}=\frac{\delta ^{2}E}{\delta
R_{kk^{\prime }}^{\ast }\delta R_{ll^{\prime }}^{\ast }}.  \label{AB}
\end{equation}%
If the two-body Hamiltonian is density independent the matrices $A$ and $B$
have the simple forms \cite{Rin.80}%
\begin{eqnarray}
A_{kk^{\prime },ll^{\prime }} &=&\left\langle \Phi \right\vert [\alpha
_{k^{\prime }}^{{}}\alpha _{k}^{{}},[\hat{H},\alpha _{l}^{+}\alpha
_{l^{\prime }}^{+}]]\left\vert \Phi \right\rangle  \nonumber \\
B_{kk^{\prime },ll^{\prime }} &=&-\left\langle \Phi \right\vert [\alpha
_{k^{\prime }}\alpha _{k},[\hat{H},\alpha _{l^{\prime }}\alpha
_{l}]]\left\vert \Phi \right\rangle .
\end{eqnarray}%
Using the representation of the Hamiltonian in the quasiparticle basis%
\begin{eqnarray}
\hat{H} &=&E_{0}+\sum_{kk^{\prime }}H_{kk^{\prime }}^{11}\alpha
_{k}^{+}\alpha _{k^{\prime }}^{{}}+\frac{1}{4}\sum_{kk^{\prime }ll^{\prime
}}H_{kk^{\prime }ll^{\prime }}^{22}\alpha _{k}^{+}\alpha _{k^{\prime
}}^{+}\alpha _{l^{\prime }}^{{}}\alpha _{l}^{{}}  \nonumber \\
&+&\sum_{kk^{\prime }ll^{\prime }}\left( H_{kk^{\prime }ll^{\prime
}}^{40}\alpha _{k}^{+}\alpha _{k^{\prime }}^{+}\alpha _{l^{\prime
}}^{+}\alpha _{l}^{+}+h.c.\right) +\sum_{kk^{\prime }ll^{\prime }}\left(
H_{kk^{\prime }ll^{\prime }}^{31}\alpha _{k}^{+}\alpha _{k^{\prime
}}^{+}\alpha _{l^{\prime }}^{+}\alpha _{l}^{{}}+h.c.\right)
\end{eqnarray}%
we find%
\begin{eqnarray}
A_{kk^{\prime }ll^{\prime }} &=&H_{kl}^{11}\delta _{k^{\prime }l^{\prime
}}-H_{k^{\prime }l}^{11}\delta _{kl^{\prime }}-H_{kl^{\prime }}^{11}\delta
_{k^{\prime }l}+H_{k^{\prime }l^{\prime }}^{11}\delta _{kl}+H_{kk^{\prime
}ll^{\prime }}^{22}  \nonumber \\
B_{kk^{\prime }ll^{\prime }} &=&4H_{kk^{\prime }ll^{\prime }}^{40}.
\end{eqnarray}%
In the quasiparticle representation the matrix $H^{11}$ is diagonal, i.e. $%
H_{kl}^{11}=E_{k}\delta _{kl}$. The matrices $H^{22}$ and $H^{40}$ are
rather complicated expressions containing the two-body $ph$- and $pp$-matrix
elements and the coefficients $U$ and $V$ (for details see Ref. \cite{Rin.80}%
)\newline
In the more general case of a density dependent Hamiltonian the same
expressions can be used, but one has to take into account the rearrangement
terms originating from the variation of the interaction with respect to the
density $\hat{\rho}$.

\bigskip


\subsection{The Relativistic QRPA in the canonical basis}

The full RQRPA equations are rather complicated,
because they require the evaluation of the matrix elements $%
H_{kk^{\prime }ll^{\prime }}^{22}$ and $H_{kk^{\prime }ll^{\prime }}^{40}$
in the basis of the Hartree-Bogoliubov spinors 
$U_{k}(\mathbf{r})$ and $V_{k}(\mathbf{r})$. 
It is considerably simpler to solve these equations in the
canonical basis, in which the relativistic 
Hartree-Bogoliubov wave functions can be
expressed in the form of BCS-like wave functions. 
In this case one needs only the matrix
elements $V_{\kappa ^{{}}\lambda ^{\prime }\kappa ^{\prime }\lambda
^{{}}}^{ph}$ of the residual 
$ph$-interaction, and the matrix elements $V_{\kappa
^{{}}\kappa ^{\prime }\lambda ^{{}}\lambda ^{\prime }}^{pp}$ 
of the pairing $pp$-interaction, as well as certain combinations of the 
occupation factors $u_{\kappa }$%
, $v_{\kappa }$. The numerical details are described in the
Appendix. In the following we use the indices $\kappa $, $\lambda $, $%
\kappa ^{\prime }$ and $\lambda ^{\prime }$ to denote states in the 
canonical basis. We emphasize that the solution 
of the relativistic quasi-particle RPA
equations in the canonical basis does not represent an approximation. 
We obtain a full solution and the results do not depend on this special 
choice of the basis.

Taking into account the rotational invariance of the nuclear system, the
quasiparticle pairs can be coupled to good angular momentum and the matrix
equations of the relativistic quasiparticle random phase approximation
(RQRPA) read 
\begin{equation}
\left(
\begin{array}{cc}
A^{J} & B^{J} \\
B^{^{\ast }J} & A^{^{\ast }J}%
\end{array}%
\right) \left(
\begin{array}{c}
X^{\nu ,JM} \\
Y^{\nu ,JM}%
\end{array}%
\right) =\omega _{\nu }\left(
\begin{array}{cc}
1 & 0 \\
0 & -1%
\end{array}%
\right) \left(
\begin{array}{c}
X^{\nu ,JM} \\
Y^{\nu ,JM}%
\end{array}%
\right) \;.  \label{rrpaeq}
\end{equation}%
For each RQRPA energy $\omega _{\nu }$, $X^{\nu }$ and $Y^{\nu }$ denote the
corresponding forward- and backward-going two-quasiparticle amplitudes,
respectively. The coupled RQRPA matrices in the canonical basis read
\begin{eqnarray}
A_{\kappa ^{{}}\kappa ^{\prime }\lambda ^{{}}\lambda ^{\prime }}^{J}
&=&H_{\kappa ^{{}}\lambda ^{{}}}^{11(J)}\delta _{\kappa ^{\prime }\lambda
^{\prime }}-H_{\kappa ^{\prime }\lambda ^{{}}}^{11(J)}\delta _{\kappa
^{{}}\lambda ^{\prime }}-H_{\kappa ^{{}}\lambda ^{\prime }}^{11(J)}\delta
_{\kappa ^{\prime }\lambda ^{{}}}+H_{\kappa ^{\prime }\lambda ^{\prime
}}^{11(J)}\delta _{\kappa ^{{}}\lambda ^{{}}}  \nonumber \\
&&+\frac{1}{2}(\xi _{\kappa ^{{}}\kappa ^{\prime }}^{+}\xi _{\lambda
^{{}}\lambda ^{\prime }}^{+}+\xi _{\kappa ^{{}}\kappa ^{\prime }}^{-}\xi
_{\lambda ^{{}}\lambda ^{\prime }}^{-})V_{\kappa ^{{}}\kappa ^{\prime
}\lambda ^{{}}\lambda ^{\prime }}^{ppJ}  \nonumber \\
&&+\zeta _{\kappa ^{{}}\kappa ^{\prime }\lambda ^{{}}\lambda ^{\prime
}}V_{\kappa \lambda^{\prime }\kappa^{\prime }\lambda}^{phJ} \\
B_{\kappa ^{{}}\kappa ^{\prime }\lambda ^{{}}\lambda ^{\prime }}^{J} &=&%
\frac{1}{2}(\xi _{\kappa ^{{}}\kappa ^{\prime }}^{+}\xi _{\lambda
^{{}}\lambda ^{\prime }}^{+}-\xi _{\kappa ^{{}}\kappa ^{\prime }}^{-}\xi
_{\lambda ^{{}}\lambda ^{\prime }}^{-})V_{\kappa ^{{}}\kappa ^{\prime
}\lambda ^{{}}\lambda ^{\prime }}^{ppJ}  \nonumber \\
&&+\zeta _{\kappa ^{{}}\kappa ^{\prime }\lambda ^{{}}\lambda ^{\prime
}}(-1)^{j_{\lambda ^{{}}}-j_{\lambda ^{\prime }}+J}V_{\kappa ^{{}}\lambda
^{{}}\kappa ^{\prime }\lambda ^{\prime }}^{phJ}\;.
\end{eqnarray}%
$H^{11}$ denotes the one-quasiparticle terms
\begin{equation}
H_{\kappa \lambda }^{11}=(u_{\kappa }u_{\lambda }-v_{\kappa }v_{\lambda
})h_{\kappa \lambda }-(u_{\kappa }v_{\lambda }+v_{\kappa }u_{\lambda
})\Delta _{\kappa \lambda }\;,
\label{H11}
\end{equation}
i.e. the canonical RHB basis does not diagonalize the Dirac single-nucleon
mean-field Hamiltonian $\hat{h}_{D}$ and the pairing field $\hat{\Delta}$.
The occupation amplitudes $v_{k}$ of the canonical states are eigenvalues of
the density matrix. $V^{ph}$ and $V^{pp}$ are the particle-hole and
particle-particle residual interactions, respectively. Their matrix elements
are multiplied by the pairing factors $\xi ^{\pm }$ and $\zeta $, defined
below by the occupation amplitudes of the canonical states. The relativistic
particle-hole interaction $V^{ph}$ is defined by the same effective
Lagrangian density as the mean-field Dirac single-nucleon Hamiltonian $\hat{h%
}_{D}$. $V^{ph}$ includes the exchange of the isoscalar scalar $\sigma $%
-meson, the isoscalar vector $\omega $-meson, the isovector vector $\rho $%
-meson, and the electromagnetic interaction. The two-body matrix elements
include contributions from the spatial components of the vector fields.
\[
\zeta _{\kappa ^{{}}\kappa ^{\prime }\lambda ^{{}}\lambda ^{\prime
}}=\left\{
\begin{array}{ll}
\eta _{\kappa ^{{}}\kappa ^{\prime }}^{+}\eta _{\lambda ^{{}}\lambda
^{\prime }}^{+} & \text{for $\sigma $, and the time-components $\omega ^{0}$%
, $\rho ^{0}$, $A^{0}$ if J is even} \\
& \text{for the space-components ${\bm\omega }$, ${\bm\rho }$, ${\bm A}$ if
J is odd} \\
\eta _{\kappa ^{{}}\kappa ^{\prime }}^{-}\eta _{\lambda ^{{}}\lambda
^{\prime }}^{-} & \text{for $\sigma $, and the time-components $\omega ^{0}$%
, $\rho ^{0}$, $A^{0}$ if J is odd} \\
& \text{for the space components ${\bm\omega }$, ${\bm\rho }$, ${\bm A}$ if
J is even} \\
&
\end{array}%
\right.
\]%
with the $\eta $-coefficients defined by
\[
\eta _{\kappa ^{{}}\kappa ^{\prime }}^{\pm }=u_{\kappa ^{{}}}v_{\kappa
^{\prime }}\pm v_{\kappa ^{{}}}u_{\kappa ^{\prime }}\;,
\]%
and
\[
\xi _{\kappa ^{{}}\kappa ^{\prime }}^{\pm }=u_{\kappa ^{{}}}u_{\kappa
^{\prime }}\mp v_{\kappa ^{{}}}v_{\kappa ^{\prime }}.
\]

The RQRPA configuration space includes the Dirac sea of negative energy
states. In addition to the configurations built from two-quasiparticle
states of positive energy, the RQRPA configuration space must also contain
pair-configurations formed from the fully or partially occupied states of
positive energy and the empty negative-energy states from the Dirac sea. The
inclusion of configurations built from occupied positive-energy states and
empty negative-energy states is essential for current conservation and the
decoupling of spurious states~\cite{Daw.90}. In recent applications of the
relativistic RPA it has been shown that the fully consistent inclusion of
the Dirac sea of negative energy states in the RRPA configuration space is
essential for a quantitative comparison with the experimental excitation
energies of giant resonances~\cite{Vre.00,Ring.01}.

It should be emphasized that the present RQRPA model is fully consistent:
the same interactions, both in the particle-hole and particle-particle
channels, are used in the RHB equation (\ref{eqhb}) that determines the
canonical quasiparticle basis, and in the RQRPA equation (\ref{rrpaeq}). In
both channels the same strength parameters of the interactions are used in
the RHB and RQRPA calculations. No additional adjustment of the parameters
is needed in RQRPA calculations. This is an essential feature of our
calculations and it ensures that RQRPA amplitudes do not contain spurious
components associated with the mixing of the nucleon number in the RHB
ground state (for $0^{+}$ excitations), or with the center-of-mass
translational motion (for $1^{-}$ excitations). 

In the next section we present results of illustrative RQRPA calculations of
the multipole response in spherical nuclei. For the multipole operator $\hat{%
Q}_{\lambda \mu }$ the response function $R(E)$ is defined
\begin{equation}
R(E,J)=\sum_{\nu }~B(J,\omega _{\nu })~\frac{1}{\pi }~\frac{\Gamma /2}{%
(E-\omega _{\nu })^{2}+(\Gamma /2)^{2}},  \label{Lorentz}
\end{equation}%
where $\Gamma $ is the width of the Lorentzian distribution, and
\begin{eqnarray}
B(J,\omega _{\nu }) &=&\bigg\vert\sum_{\kappa \kappa ^{\prime }}\bigg\{%
X_{\kappa \kappa ^{\prime }}^{\nu ,J0}\langle \kappa ||\hat{Q}_{J}||\kappa
^{\prime }\rangle  \nonumber \\
&+&~(-1)^{j_{\kappa ^{{}}}-j_{\kappa ^{\prime }}+J}\,Y_{\kappa \kappa
^{\prime }}^{\nu ,J0}\,\langle \kappa ^{\prime }||\hat{Q}_{J}||\kappa
\rangle \,\bigg\}(u_{\kappa ^{{}}}v_{\kappa ^{\prime }}+~(-1)^{J}v_{\kappa
^{{}}}u_{\kappa ^{\prime }})\bigg\vert^{2}.  \label{strength}
\end{eqnarray}%
In all the examples considered in Sec.~\ref{secIII}, the discrete strength
distribution are folded by a Lorentzian of width $\Gamma $=1MeV . For the
state $\left\vert J,\nu \right\rangle $, the RQRPA transition density reads
\begin{eqnarray}
\delta \rho _{J}^{\nu }(r) &=&\sum_{\kappa \kappa ^{\prime }}\bigg\{\langle
\kappa ||Y_{J}||\kappa ^{\prime }\rangle f_{\kappa ^{{}}}(r)f_{\kappa
^{\prime }}(r)+\langle \hat{\kappa}||Y_{J}||\hat{\kappa}^{\prime }\rangle
g_{\kappa ^{{}}}(r)g_{\kappa ^{\prime }}(r)\bigg\}  \nonumber \\
&&\cdot \Big(X_{\kappa \kappa ^{\prime }}^{\nu ,J0}+(-1)^{J}Y_{\kappa \kappa
^{\prime }}^{\nu ,J0}\Big)(u_{\kappa ^{{}}}v_{\kappa ^{\prime
}}+~(-1)^{J}v_{\kappa ^{{}}}u_{\kappa ^{\prime }}),  \label{trdens}
\end{eqnarray}%
where $\kappa $ and $\hat{\kappa}$ denote the quantum numbers of the large
and small components of the Dirac spinors, respectively. $f_{\kappa }(r)$
and $g_{\kappa }(r)$ are the corresponding large and small radial components.


\section{\label{secIII}Illustrative calculations and tests of the RQRPA}


Nuclear properties calculated with the RHB+RQRPA model will, of course,
crucially depend on the choice of the effective RMF Lagrangian in the $ph$%
-channel, as well as on the treatment of pairing correlations. The most
successful RMF effective interactions are purely phenomenological, with
parameters adjusted to reproduce the nuclear matter equation of state and a
set of global properties of spherical closed-shell nuclei. In most
applications of the RHB model, in particular, we have used the NL3 effective
interaction \cite{Lal.97} for the RMF effective Lagrangian. Properties
calculated with NL3 indicate that this is probably the best nonlinear
effective interaction so far, both for nuclei at and away from the line of $%
\beta $-stability. In the $pp$-channel of the RHB model we have used a
phenomenological pairing interaction, the pairing part of the Gogny force,
\begin{equation}
V^{pp}(1,2)~=~\sum_{i=1,2}e^{-((\mathbf{r}_{1}-\mathbf{r}_{2})/{\mu _{i}}%
)^{2}}\,(W_{i}~+~B_{i}P^{\sigma }-H_{i}P^{\tau }-M_{i}P^{\sigma }P^{\tau }),
\end{equation}
with the set D1S \cite{Ber.84} for the parameters $\mu _{i}$, $W_{i}$, $%
B_{i} $, $H_{i}$ and $M_{i}$ $(i=1,2)$. This force has been very carefully
adjusted to the pairing properties of finite nuclei all over the periodic
table. In particular, the basic advantage of the Gogny force is the finite
range, which automatically guarantees a proper cut-off in momentum space.
All RHB+RQRPA calculations presented in this work have been performed with
the NL3+D1S combination of effective interactions.

In order to illustrate the RHB+RQRPA approach and to test the numerical
implementation of the RQRPA equations, in this section we calculate the
isoscalar monopole, isovector dipole and isoscalar quadrupole response of $%
^{22}$O. Similar calculations for the neutron-rich oxygen isotopes were
recently performed by Matsuo~\cite{Mat.01,Mat.02} in the framework of the
non-relativistic continuum linear response theory based on the
Hartree-Fock-Bogoliubov formalism in coordinate state representation. The
two theoretical frameworks differ, of course, both in the physical contents,
as well as in the numerical implementation. The results can, nevertheless,
be compared at least at the qualitative level. In the HFB+QRPA model of
Refs.~\cite{Mat.01,Mat.02} a Woods-Saxon parameterization is adopted for the
single-particle potential, and a Skyrme-type density dependent delta force
is used for the residual interaction in the $ph$-channel of the QRPA. Since
the calculation of the single-particle potential and $ph$-interaction is not
self-consistent, the interaction strength of the residual interaction is
renormalized for each nucleus in such a way that the dipole response has a
zero-energy mode corresponding to the spurious center of mass motion. For
the pairing interaction, a density-dependent delta force is used both in the
calculation of the HFB pairing field for the ground state, and in the linear
response equation for the excitations. The calculation is consistent in the $%
pp$-channel. The present RHB+RQRPA calculations are fully self-consistent:
the same combination of effective interactions, NL3 in the $ph$-channel and
Gogny D1S in the $pp$-channel, are used both in the RHB calculation of the
ground state and as RQRPA residual interactions. The parameters of the RQRPA
residual interactions have exactly the same values as those used in the RHB
calculation.

In the analysis of Refs.~\cite{Mat.01,Mat.02}, Matsuo has illustrated the
importance of a consistent treatment of pairing correlations in the HFB+QRPA
framework. The residual pairing interaction in the QRPA generates pronounced
dynamical correlation effects on the responses through pair density
fluctuations. Moreover, the energy weighted sum rules are only satisfied if
the pairing interaction is consistently included both in the static HFB and
in the dynamical linear response. We have verified that the results obtained
in the HFB+QRPA framework are also reproduced in the RHB+RQRPA calculations.

In the left panel of Fig.~\ref{fig1} we display the monopole strength
function of the neutron number operator in $^{22}$O. There should be no
response to the number operator since it is a conserved quantity, i.e. the
Nambu-Goldstone mode associated with the nucleon number conservation should
have zero excitation energy. The dashed curve (no dynamical pairing)
represents the strength function obtained when the pairing interaction is
included only in the RHB calculation of the ground state, but not in the
residual interaction of the RQRPA. The solid line (zero response)
corresponds to the full RHB+RQRPA calculation, with the pairing interaction
included both in the RHB ground state, and in the RQRPA residual
interaction. The same result was also obtained in the HFB+QRPA calculation
for $^{24}$O in Ref.~\cite{Mat.02}: the spurious strength of the number
operator appears when the pairing interaction is included only in the
stationary solution for the ground state, i.e. when the dynamical QRPA
pairing correlations are neglected.

The isoscalar strength functions of the monopole operator $\sum^{A}_{i=1}
r^2_{i}$ in $^{22}$O, shown in the right panel of Fig.~\ref{fig1},
correspond to three different calculations: a) the RMF+RRPA calculation
without pairing, b) pairing correlations are included in the RHB calculation
of the ground state, but not in the RQRPA residual interaction (no dynamical
pairing), and c) the fully self-consistent RHB+RQRPA calculation. Just as in
the case of the number operator, by including pairing correlations only in
the RHB ground state a strong spurious response is generated below 10 MeV.
The Nambu-Goldstone mode is found at zero excitation energy (in this
particular calculation it was located below 0.2 MeV) only when pairing
correlations are consistently included also in the residual RQRPA
interaction. When the result of the full RHB+RQRPA is compared with the
response calculated without pairing, one notices that, as expected, pairing
correlations have relatively little influence on the response in the region
of giant resonances above 20 MeV. A more pronounced effect is found at lower
energies. The fragmentation of the single peak at $\approx$ 12.5 MeV
reflects the broadening of the Fermi surface by the pairing correlations.

The isovector strength function ($J^{\pi}=1^{-}$) of the dipole operator
\begin{equation}
\hat{Q}_{1 m}^{T=1} \ = \frac{N}{N+Z}\sum^{Z}_{p=1} r_{p}Y_{1 m} - \frac{Z}{%
N+Z}\sum^{N}_{n=1} r_{n}Y_{1 m}  \label{dipop}
\end{equation}
for $^{22}$O is displayed in the left panel of Fig.~\ref{fig2}. In this
example we also compare the results of the RMF+RRPA calculations without
pairing, with pairing correlations included only in the RHB ground state (no
dynamical pairing), and with the fully self-consistent RHB+RQRPA response. A
large configuration space enables the separation of the zero-energy mode
that corresponds to the spurious center of mass motion. In the present
calculation for $^{22}$O this mode is found at $E = 0.04$ MeV.

The isovector dipole response in neutron-rich oxygen isotopes has recently
attracted considerable interest because these nuclei might be good
candidates for a possible identification of the low-lying collective soft
mode (pygmy state), that corresponds to the oscillations of excess neutrons
out of phase with the core composed of an equal number of protons and
neutrons~\cite{Try.01,Lei.01}.
The strength functions shown in Fig.~\ref{fig2} illustrate the importance of
including pairing correlations in the calculation of the isovector dipole
response. Pairing is, of course, particularly important for the low-lying
strength below 10 MeV. The inclusion of pairing correlations in the full
RHB+RQRPA calculation enhances the low-energy dipole strength near the
threshold. For the main peak in the low-energy region ($\approx$ 8.65 MeV),
in the right panel of Fig.~\ref{fig2} we display the proton and neutron
transition densities. In contrast to the well known radial dependence of the
IVGDR transition densities (proton and neutron densities oscillate with
opposite phases, the amplitude of the isovector transition density is much
larger than that of the isoscalar component), the proton and neutron
transition densities for the main low-energy peak are in phase in the
nuclear interior, there is no contribution from the protons in the surface
region, the isoscalar transition density dominates over the isovector one in
the interior, and the strong neutron transition density displays a long tail
in the radial coordinate. A similar behavior has been predicted for the
light neutron halo nuclei $^{6}$He, $^{11}$Li and $^{12}$Be in Ref.~\cite%
{Sag.96}, where it has been shown that the long tails of the wave functions
of the loosely-bound neutrons are responsible for the different radial
dependence of the transition densities that correspond to the soft
low-energy states as compared to those of the giant resonances.

The effect of pairing correlations on the isovector dipole response in $%
^{22} $O is very similar to the one obtained in the HFB+QRPA framework (Fig.
8 of Ref.~\cite{Mat.02}). In the low-energy region below 10 MeV, however,
the pairing interaction used in the QRPA calculation produces a much
stronger enhancement of the dipole strength, as compared to the results
shown in Fig.~\ref{fig2}. The reason probably lies in the choice of the
pairing interaction. While we use the volume Gogny pairing, in Ref.~\cite%
{Mat.02} a density-dependent delta force was used in the $pp$ channel. This
interaction is surface peaked and therefore produces a stronger effect on
the low-energy dipole strength near the threshold. Nevertheless, we
emphasize that the RHB+RQRPA results for the low-lying dipole strength
distribution in $^{22}$O are in very good agreement with recent experimental
data~\cite{Lei.01}.

In the left panel of Fig.~\ref{fig3} we display the RHB+RQRPA isoscalar and
isovector quadrupole ($J^{\pi}=2^{+}$) strength distributions in $^{22}$O.
The low-lying $J^{\pi}=2^{+}$ state is calculated at $E = 2.95$ MeV, and
this value should be compared with the experimental excitation energy of the
first $2^{+}$ state: 3.2 MeV~\cite{Bell.01}. The strong peak at $E = 22.3$
MeV in the isoscalar strength function corresponds to the isoscalar giant
quadrupole resonance (IS GQR). The isovector response, on the other hand, is
strongly fragmented over the large region of excitation energies $E\simeq
18-38$ MeV. The effect of pairing correlations on the isoscalar response is
illustrated in the right panel of Fig.~\ref{fig3}, where again the full
RHB+RQRPA strength function is compared to the RMF+RRPA calculation without
pairing, and with the response obtained when the pairing interaction is
included only in the RHB ground state (no dynamical pairing). As one would
expect, the effect of pairing correlations is not particularly pronounced in
the giant resonance region. The inclusion of pairing correlations, however,
has a relatively strong effect on the low-lying $2^{+}$ state. This is seen
more clearly in the left panel of Fig.~\ref{fig4}, where only the low-energy
portion of the isoscalar strength distributions in $^{22}$O is shown. With
respect to the RRPA calculation, the inclusion of the pairing interaction in
the static solution for the ground state increases the excitation energy of
the lowest $2^{+}$ state by $\approx$ 3 MeV. The fully self-consistent
RHB+RQRPA calculation lowers the excitation energy from $\approx$ 4.5 MeV to
$E = 2.95$ MeV. The inclusion of pairing correlations increases the
collectivity of the low-lying $2^{+}$ state. A very similar result for the
low-lying quadrupole state in $^{24}$O has been obtained by Matsuo in the
HFB+QRPA framework~\cite{Mat.02}. The proton and neutron transition
densities for the $2^{+}$ state at $E = 2.95$ MeV are shown in the right
panel of Fig.~\ref{fig4}. They display a characteristic radial dependence.
Both transition densities are, of course, peaked in the surface region, but
the proton contribution is much smaller. The RHB+RQRPA results for the $%
2^{+} $ excitations are in agreement with non-relativistic QRPA calculations
of the quadrupole response in neutron rich oxygen isotopes~\cite%
{Kha.00,Kha.00e,Mat.02,Kha.02}.


\section{\label{secIV}Evolution of the low-lying isovector dipole strength
in Sn isotopes and N=82 isotones}


The dipole response of very neutron-rich isotopes is characterized by the
fragmentation of the strength distribution and its spreading into the
low-energy region, and by the mixing of isoscalar and isovector modes. It
appears that in most relatively light nuclei the onset of dipole strength in
the low-energy region is due to non-resonant independent single particle
excitations of the loosely bound neutrons. The structure of the low-lying
dipole strength changes with mass. As we have shown in the RRPA analysis of
Ref.~\cite{Vrepyg2.01}, in heavier nuclei low-lying dipole states appear
that are characterized by a more distributed structure of the RRPA
amplitude. Among several peaks characterized by single particle transitions,
a single collective dipole state is identified below 10 MeV, and its
amplitude represents a coherent superposition of many neutron particle-hole
configurations.

Very recently experimental data have been reported on the concentration of
electric dipole strength below the neutron separation energy in $N=82$
semi-magic nuclei. The distribution of the electric dipole strength in $%
^{138}$Ba, $^{140}$Ce, and $^{144}$Sm displays a resonant structure between
5.5 MeV and 8 MeV, exhausting $\approx$ 1\% of the isovector E1 EWSR~\cite%
{Zil.02}. In $^{138}$Ba negative parity quantum numbers have been assigned
to 18 dipole excitations between 5.5 MeV and 6.5 MeV~\cite{Pie.02}.

In Figs.~\ref{fig5} and \ref{fig6} we display the isovector dipole strength
distributions in eight $N=82$ isotones, calculated in the RHB+RQRPA
framework with the NL3+D1S combination of effective interactions. The
calculation is fully self-consistent, with the Gogny finite-range pairing
included both in the RHB ground state, and in the RQRPA residual
interaction. The isovector dipole response is shown for even-Z nuclei from $%
^{146}$Gd to the doubly magic $^{132}$Sn. In addition to the characteristic
peak of the isovector giant dipole resonance (IVGDR) at $\approx$ 15 MeV,
the evolution of the low-lying dipole strength with decreasing proton number
is clearly observed below 10 MeV. The strength of the low-lying dipole
response increases with the relative increase of the neutron contribution,
i.e. with reducing the number of protons. For the main peaks in the
low-energy region below 10 MeV, in the panels on the right side of Figs.~\ref%
{fig5} and \ref{fig6} we display the corresponding neutron and proton
transition densities. The radial dependence is very different from that of
the transition densities of the IVGDR peak. For all eight nuclei the main
peak below 10 MeV does not correspond to an isovector excitation, i.e. the
proton and neutron transition densities have the same sign. The relative
contribution of the protons in the surface region decreases with reducing
the proton number. In particular, for the nuclei shown in Fig.~\ref{fig6}: $%
^{138}$Ba, $^{136}$Xe, $^{134}$Te and $^{132}$Sn, there is practically no
proton contribution to the transition density beyond 6 fm. The dynamics is
that of a pygmy resonance: the neutron skin oscillates against the core. In
Ref.~\cite{Zil.02} it was emphasized that the observed low-lying dipole
states in the N=82 isotones are not just statistical E1 excitations sitting
on the tail of the GDR, but represent a fundamental structure effect. In
Fig.~\ref{fig7} we show that this is also the case for the RHB+RQRPA
results. For the dipole strength distribution of $^{140}$Ce, shown in the
left panel, in the right column we compare the neutron and proton transition
densities for the IVGDR peak at 14.31 MeV, for the peak at 12.51 MeV, and
for the main peak in the low-energy region at 8.22 MeV. The peak at 12.51
MeV, as well as other peaks in the interval 10-14 MeV, displays transition
densities very similar to those of the GDR peak, i.e. these states belong to
the tail of the GDR. The dynamics of the low-energy mode at 8.22 MeV, on the
other hand, is very different: the proton and neutron transition densities
are in phase in the nuclear interior, there is almost no contribution from
the protons in the surface region, the isoscalar transition density
dominates over the isovector one, and the peak of the strong neutron
transition density in the surface region is shifted toward larger radii.

On a quantitative level, the present RHB+RQRPA calculation does not compare
too well with the experimental data on the low-lying dipole strength in the
N=82 isotones. First, while the observed low-energy dipole states in $^{138}$%
Ba, $^{140}$Ce, and $^{144}$Sm are concentrated between 5.5 MeV and 8 MeV,
the calculated pygmy states in these nuclei are above 8 MeV. This can be
partly explained by the low effective nucleon mass of the NL3 mean-field
interaction~\cite{VNR.02}. On the other hand, the excitation energies of the
IV GDR are, as will be shown below in the example of Sn isotopes, rather
well reproduced by the NL3 interaction. The fact that NL3 reproduces the IV
GDR, but not the centroid of the low-energy dipole strength, might indicate
that the isovector channel of this force needs a better parameterization.
Second and more important, the number of RQRPA peaks below 10 MeV, for the
operator (\ref{dipop}), is much smaller than the number of observed dipole
states in the low-energy region~\cite{Pie.02,Zil.02}. The observed low-lying
E1 strength consists of many states of different origin. This has been
discussed in Ref.~\cite{Zil.02}. In addition to the two-phonon and
three-phonon states, and the soft pygmy state, in this energy region one
could also expect some compressional low-lying isoscalar dipole strength~%
\cite{CLY.01}, maybe mixed with toroidal states~\cite{Vre.02,Rye.02}, as well as
the E1 strength generated by the breaking of the isospin symmetry due to a
clustering mechanism~\cite{Iac.85}. A detailed investigation of the nature
of all observed low-lying dipole states in N=82 nuclei is, of course, beyond
the scope of the present analysis, since our model space does not include
multi-phonon configurations.

The Sn isotopes present another very interesting example of the evolution of
the low-lying dipole strength with neutron number~\cite{Gov.98}. 
In Ref.~\cite{Vrepyg2.01}
we have performed an analysis of the isovector dipole response of
neutron-rich Sn isotopes in the relativistic RPA framework. The RMF+RRPA
calculation has shown that, among several dipole states in the low-energy
region between 7 MeV and 9 MeV, and characterized by single particle
transitions, a single state is found with a more distributed structure of
the RRPA amplitude, exhausting approximately 2\% of the EWSR. The results of
the fully self-consistent RHB+RQRPA calculation, with the NL3+D1S
combination of effective interactions, are shown in Figs.~\ref{fig8} and \ref%
{fig9}: the isovector dipole strength functions of the Sn isotopes (left
panels), and the corresponding proton and neutron transition densities for
the main peaks in the low-energy region (right panels). With the increase of
the number of neutrons a relatively strong peak appears below 10 MeV,
characterized by the dynamics of the pygmy resonance (see the transition
densities). The low-energy pygmy peak is most pronounced in $^{124}$Sn. It
does not become stronger by further increasing the neutron number, and
additional fragmentation of the low-lying strength is observed in $^{132}$%
Sn. For the Sn isotopes we can compare the RHB+RQRPA results with
experimental data on IV GDR. In the upper panel of Fig.~\ref{fig10} the
experimental IVGDR excitation energies~\cite{Ber.75} are shown in comparison
with the calculated $E_{\mathrm{GDR}}$. The energy of the resonance is
defined as the centroid energy 
\begin{equation}  \label{meanen}
\bar E = \frac{m_1}{m_0}~,
\end{equation}
with the energy weighted moments for discrete spectra
\begin{equation}
m_k=\sum_{\nu}B(J,\omega_\nu) E^k_{\nu}.  \label{m1}
\end{equation}
For $k=1$ this equation defines the energy weighted sum rule (EWSR). The
calculated energies of the IV GDR are in excellent agreement with
experimental data, and the mass dependence of the excitation energies is
reproduced in detail. In the middle panel of Fig.~\ref{fig10} we plot the
calculated energies of the pygmy states. In comparison with the IV GDR, the
excitation energies of the pygmy states decrease more steeply with the mass
number. The ratio of the energy weighted m$_1$ moments calculated in the low
(E$\leq$10 MeV) and high (E$>$10 MeV) energy regions, as function of the
mass number, is plotted in the lower panel of Fig.~\ref{fig10}. The relative
contribution of the low-energy region increases with the neutron excess. The
ratio m$_{1,\mathrm{LOW}}$/ m$_{1,\mathrm{HIGH}}$ reaches a maximum $\approx
0.06$ for $^{124}$Sn, and it slowly decreases to $\approx 0.05$ for $^{132}$%
Sn. 


\section{\label{secV}Summary}


In this work we have formulated the relativistic QRPA in the canonical
single-nucleon basis of the relativistic Hartree-Bogoliubov (RHB) model. The
RHB model presents the relativistic extension of the Hartree-Fock-Bogoliubov
framework, and it provides a unified description of mean-field and pairing
correlations. A consistent and unified treatment of the $ph$ and $pp$
channels is very important for weakly bound nuclei far from stability. In
the RHB framework the ground state of a nucleus can be written either in the
quasiparticle basis as a product of independent quasi-particle states, or in
the canonical basis as a highly correlated BCS-state. By definition, the
canonical basis diagonalizes the density matrix and it is always localized.
It describes both the bound states and the positive-energy single-particle
continuum. The QRPA model employed in this work is fully self-consistent.
For the interaction in the particle-hole channel effective Lagrangians with
nonlinear meson self-interactions are used, and pairing correlations are
described by the pairing part of the finite range Gogny interaction. Both in
the $ph$ and $pp$ channels, the same interactions are used in the RHB
equations that determine the canonical quasiparticle basis, and in the
matrix equations of the RQRPA. This is very important, because the energy
weighted sum rules are only satisfied if the pairing interaction is
consistently included both in the static RHB and in the dynamical RQRPA
calculations. The two-quasiparticle configuration space includes states with
both nucleons in the discrete bound levels, states with one nucleon in the
bound levels and one nucleon in the continuum, and also states with both
nucleons in the continuum. The RQRPA configuration space includes the Dirac
sea of negative energy states. In addition to the configurations built from
two-quasiparticle states of positive energy, the RQRPA configuration space
contains pair-configurations formed from the fully or partially occupied
states of positive energy and the empty negative-energy states from the
Dirac sea. The inclusion of configurations built from occupied
positive-energy states and empty negative-energy states is essential for the
decoupling of spurious states.

The RHB+RQRPA approach has been tested in the example of the isoscalar
monopole, isovector dipole and isoscalar quadrupole excitations of $^{22}$O.
The NL3 parameterization has been used for the RMF effective Lagrangian, and
the Gogny D1S finite range interaction has been employed in the $pp$
channel. In the present numerical implementation the RHB eigenvalue
equations, the Klein-Gordon equations for the meson fields, and the RQRPA
matrix equations are solved by expanding the nucleon spinors and the meson
fields in a basis of eigenfunctions of a spherical harmonic oscillator. The
calculations have illustrated the importance of a consistent treatment of
pairing correlations in the RHB+RQRPA framework. The results have been
compared with calculations performed in the non-relativistic continuum QRPA
based on the coordinate state representation of the HFB framework. It has
been shown that the RHB+RQRPA results are in agreement with recent
experimental data and with non-relativistic QRPA calculations of the
multipole response of neutron rich oxygen isotopes.

The RHB+RQRPA has been employed in the analysis of the evolution of the
low-lying isovector dipole strength in Sn isotopes and N=82 isotones. The
analysis is motivated by very recent data on the concentration of electric
dipole strength below the neutron separation energy in $N=82$ semi-magic
nuclei. It has been shown that in neutron rich nuclei a relatively strong
peak appears in the dipole response below 10 MeV, with a QRPA amplitude
characterized by a coherent superposition of many neutron quasiparticle
configurations. The dynamics of this state corresponds to that of a pygmy
dipole resonance: the oscillation of the skin of excess neutrons against the
core composed of an equal number of protons and neutrons. It should be
emphasized that, even though the IV GDR excitation energies calculated with
the NL3 effective interaction are in excellent agreement with experimental
data on Sn isotopes, the pygmy peaks in the low-energy region do not compare
too well with the data on low-lying dipole strength in $N=82$ isotones. The
calculated peaks are $\approx$ 2 MeV higher than the experimental weighted
mean energies. This might indicate that there are problems with the
isovector channel of the effective interaction and with the effective mass.
Namely, if the pygmy resonance is directly related to the thickness of the
neutron skin, the splitting between the excitation energies of the pygmy
state and the IV GDR should be determined by the isovector channel of the
effective force. A detailed quantitative analysis of the empirical low-lying
isovector dipole response of neutron rich $N=82$ nuclei in the RHB+RQRPA
framework will be included in a forthcoming publication.

Summarizing, the relativistic QRPA formulated in the canonical basis of the
RHB model represents a significant contribution to the theoretical tools
that can be employed in the description of the multipole response of
unstable weakly bound nuclei far from stability.

\bigskip \bigskip

\section{Appendix: Numerical details of the solution of the 
RQRPA equations in the canonical basis}

The relativistic quasi-particle RPA equations can be simplified 
considerably by employing the canonical basis. According to the
theorem of Bloch and Messiah~\cite{Blo.62}, any RHB wave function can be
expressed either in the quasiparticle basis as a product of independent
quasiparticle states, or in the canonical basis as a highly correlated
BCS-state. For systems with an even number of particles we have%
\begin{equation}
|\Phi \rangle ~=~\prod_{\kappa >0}(u_{\kappa }+v_{\kappa }a_{\kappa
}^{\dagger }a_{\overline{\kappa }}^{\dagger })|-\rangle \;.
\end{equation}%
$|-\rangle $ denotes the nucleon vacuum, the operators $a_{\kappa }^{\dagger
}$ and $a_{\bar{\kappa}}^{\dagger }$ create nucleons in the canonical basis.
The occupation
probabilities are given by
\begin{equation}
v_{\kappa }^{2}=\frac{1}{2}\left( 1-\frac{\varepsilon _{\kappa
}^{{}}-m-\lambda }{\sqrt{(\varepsilon _{\kappa }^{{}}-m-\lambda
)^{2}+\Delta _{\kappa }^{2}}}\right). \label{occup}
\end{equation}%
$\varepsilon _{\kappa }^{{}}=\left\langle \kappa |\hat{h}_{D}|\kappa
\right\rangle $ and $\Delta _{\kappa }^{{}}=$ $\left\langle \kappa |\hat{%
\Delta}|\overline{\kappa }\right\rangle $ are the diagonal elements of the
Dirac single-particle Hamiltonian and the pairing field in the canonical
basis, respectively. In contrast to the BCS framework, however, neither of
these fields is diagonal in the canonical basis. The basis itself is
specified by the requirement that it diagonalizes the single-nucleon density
matrix $\hat{\rho}(\mathbf{r,r}^{\prime })=\sum_{k>0}V_{k}^{{}}(\mathbf{r}%
)V_{k}^{\dagger }(\mathbf{r}^{\prime })$. The transformation to the
canonical basis determines the energies and occupation probabilities of
single-nucleon states, that correspond to the self-consistent solution for
the ground state of a nucleus. Since it diagonalizes the density matrix, the
canonical basis is localized. It describes both the bound states and the
positive-energy single-particle continuum~\cite{Dob.96}.

Many of the eigenvalues (\ref{occup}) of the density matrix are identically
zero. In particular, those at very high energies in the continuum, but also
those that correspond to the levels in the Dirac sea (\textit{no-sea}
approximation). Because of this degeneracy the levels in the canonical basis
are not uniquely determined by the numerical diagonalization of the density
matrix $\hat{\rho}(\mathbf{r,r}^{\prime })$. In addition to the well defined
eigenstates $\left\vert \kappa \right\rangle $ with non-degenerate
eigenvalues $0<v_{\kappa }^{2}<1$, there is one set of eigenstates with
eigenvalues equal 0 and another set of eigenstates with eigenvalues equal 1.
Any linear combination of eigenstates with eigenvalue 0 (1) is again an
eigenstate with eigenvalue 0 (1). The diagonal pairing matrix elements $%
\Delta _{\mu }^{{}}$ vanish in these degenerate subspaces. The corresponding
single particle energies $\varepsilon _{\mu }$, however, are arbitrary and
unphysical. Within these two subspaces the canonical basis is not uniquely
defined.

We therefore introduce an additional requirement, that the canonical basis
in each of these subspaces diagonalizes the single particle Hamiltonian $%
\hat{h}_{D}$. In practical applications one thus first diagonalizes the
matrix $\hat{\rho}$. This gives all the canonical basis states with $%
0<v_{\kappa }^{2}<1$, and in addition two sets of degenerate eigenstates
with eigenvalues 0 and 1. Two eigenstates $\left\vert \kappa
\right\rangle $ and  $\left\vert \lambda \right\rangle $ are considered 
degenerate if the corresponding eigenvalues differ 
less than a given parameter $%
\epsilon _{d}$:
\begin{equation}
|v_{\kappa }^{2}-v_{\lambda }^{2}|<\epsilon _{d}.  \label{epsilond}
\end{equation}%
In the second step the single particle Hamiltonian $\hat{h}_{D}$ is
diagonalized in the subspace of degenerate eigenvectors of the density
matrix with eigenvalues 0 (1). These new vectors are also eigenvectors of $%
\hat{\rho}$ with eigenvalues 0 (1). This procedure uniquely determines the
energies $\varepsilon _{\kappa }$ and occupation probabilities $v_{\kappa
}^{2}$ of single-particle states, that correspond to the self-consistent
solution for the ground state of a nucleus. An appropriate choice, 
of course, has to be made for the parameter $%
\epsilon _{d}$. If it is too large, a linear combination of the 
eigenstates $\left\vert \kappa
\right\rangle $ and  $\left\vert \lambda \right\rangle $ 
that diagonalizes $\hat{h}_{D}$, will no longer
be an eigenvector of the density matrix $\hat{\rho}$. 

It is important to illustrate how the 
RQRPA results depend on the choice of the parameter
$\epsilon _{d}$ in Eq. (\ref{epsilond}). 
For the nucleus $^{22}$O, in Fig.~\ref{fig11}(a) we
display the isovector dipole strength distributions,
calculated with $\epsilon _{d}=10^{-4}-10^{-7}$. 
For any two values of $\epsilon_{d}>10^{-6}$ the corresponding
strength distributions show pronounced differences.
When $\epsilon_{d} \leq 10^{-6}$, the dipole response does not depend
any longer on its precise numerical value, and the 
spurious Nambu-Goldstone $1^{-}$ mode is found at an
excitation energy $\leq 0.1$ MeV. 

The RQRPA matrix is diagonalized in the finite dimensional 
two-quasiparticle ($2qp$) vector space.
There are two types of $2qp$ states: 1) those built from $qp$ states
of positive energy, and 2) those formed by one fully or partially 
occupied state of positive energy and one empty negative energy state
from the Dirac sea.
The dimension of the RQRPA configuration space is thus determined by two
cut-off parameters: $E_{Cp}$ is the maximum value of
the sum of the diagonal matrix elements of  
$H^{11}$ (\ref{H11}) for the first type of $2qp$ states, 
and $E_{Ca}$ is the maximum absolute value of the
sum of the diagonal matrix elements of $H^{11}$ (\ref{H11}) for $2qp$ states with
one quasiparticle in the Dirac sea. 
The choice of the two cut-off parameters $E_{Cp}$ and $E_{Ca}$
is restricted by the following conditions:
(a) there should be no
response to the number operator, i.e. the
Nambu-Goldstone $0^{+}$ mode associated with the 
nucleon number conservation should
have zero excitation energy, 
(b) the spurious excitation corresponding to the translation of the
nucleus decouples as a zero-energy excitation mode, and 
(c) the response function does not depend on the precise numerical 
values of $E_{Cp}$ and $E_{Ca}$.

In Fig.~\ref{fig11}(b) we show
how the response to the neutron number operator for $^{22}$O varies
with the cut-off parameter $E_{Cp}$ in the range 30 -- 270 MeV. 
The choice $E_{Ca}$=1700 MeV
includes the entire negative-energy Dirac spectrum. 
The response is obviously reduced as
the number of $2qp$ configurations increases. Already for $E_{Cp}=$ 90 MeV
the Nambu-Goldstone $0^{+}$ mode converges to $\leq 0.1$ MeV. 

A large configuration space is also necessary 
in order to bring the spurious
$1^{-}$ state at zero excitation energy. 
In Fig.~\ref{fig11}(c) we illustrate the convergence of the energy of
the $1^{-}$ spurious state in $^{22}$O and $^{120}$Sn. The excitation 
energies are plotted as functions
of the energy cut-off parameter $E_{Cp}$. $E_{Ca}$ is kept at 1700 MeV. 

The choice of the cut-off parameter $E_{Ca}$ has a pronounced 
influence on the calculated isoscalar monopole response. This 
is illustrated in 
Fig.~\ref{fig11}(d), where we show how the energies of the giant 
monopole resonance (GMR) in $^{22}$O and $^{120}$Sn depend on the 
value of $E_{Ca}$. For $E_{Ca}\leq$ 1150 MeV, only positive-energy
$2qp$ states are included in the RQRPA basis 
and the excitation energies of the GMR peaks 
are simply too low. As $E_{Ca}$ is increased to include 
the negative energy states, the GMR excitation energies 
also increase and saturate for $E_{Ca}\geq$1500 MeV.
\leftline{\bf ACKNOWLEDGMENTS}

This work has been supported in part by the Bundesministerium
f\"ur Bildung und Forschung under project 06 TM 979, and by the
Gesellschaft f\" ur Schwerionenforschung (GSI) Darmstadt. T. N.
acknowledges the support from the Alexander von Humboldt -
Stiftung.

==========================================================================

\newpage

\begin{figure}
\caption{The strength function for the neutron number operator
(left), and the isoscalar strength function for the monopole
operator (right) in $^{22}$O. The curves correspond to the
RMF+RRPA calculation without pairing (dotted), with pairing
correlations included in the RHB calculation of the ground state,
but not in the RRPA residual interaction (dashed), and to the
fully self-consistent RHB+RQRPA calculation (solid).} 
\label{fig1}
\end{figure}

\begin{figure}
\caption{The isovector strength function of the dipole operator in
$^{22}$O (left). The fully self-consistent RHB+RQRPA response
(solid line) is compared with the RMF+RRPA calculation without
pairing (dotted line), and with the RHB+RRPA calculation that
includes pairing correlations only in the ground state (dashed
line). The proton and neutron transition densities for the peak at
$E=8.65$ MeV are shown in the right panel.} 
\label{fig2}
\end{figure}

\begin{figure}
\caption{The RHB+RQRPA isoscalar and isovector quadrupole strength
distributions in $^{22}$O (left panel). In the right panel the
full RHB+RQRPA isoscalar strength function (solid) is compared to
the RMF+RRPA calculation without pairing (dotted), and with the
response obtained when the pairing interaction is included only in
the RHB ground state (dashed).} 
\label{fig3}
\end{figure}

\begin{figure}
\caption{Low-energy portion of the isoscalar quadrupole strength
distribution in $^{22}$O (left). The neutron and proton transition
densities for the $J^{\pi}=2^{+}$ state at $E=2.95$ MeV (right).}
\label{fig4}
\end{figure}

\begin{figure}
\caption{RHB+RQRPA isovector dipole strength distributions in
$^{146}$Gd, $^{144}$Sm, $^{142}$Nd and $^{140}$Ce, calculated with
the NL3+D1S effective interaction. The corresponding proton and
neutron transition densities for the main peak in the low-energy
region below 10 MeV are displayed in the panels on the right
side.} \label{fig5}
\end{figure}

\begin{figure}
\caption{Same as in Fig.~\protect\ref{fig5}, but for the $N = 82$
isotones: $^{138}$Ba, $^{136}$Xe, $^{134}$Te and $^{132}$Sn.}
\label{fig6}
\end{figure}

\begin{figure}
\caption{The isovector dipole strength distribution in $^{140}$Ce
(left panel). The neutron and proton transition densities for the
IVGDR peaks at 14.31 MeV, 12.51 MeV, and for the main peak in the
low-energy region at 8.22 MeV (right).} \label{fig7}
\end{figure}

\begin{figure}
\caption{RHB+RQRPA isovector dipole strength distributions in Sn
isotopes, calculated with the NL3+D1S effective interaction. The
corresponding proton and neutron transition densities for the main
peak below the IVGDR are displayed in the panels on the right
side.} \label{fig8}
\end{figure}

\begin{figure}
\caption{Same as in Fig.~{\protect\ref{fig8}}, but for the heavier
Sn isotopes.} \label{fig9}
\end{figure}

\begin{figure}
\caption{In the upper panel the experimental IV GDR excitation
energies of the Sn isotopes are compared with the RHB+RQRPA
results calculated with the NL3+D1S effective interaction. The
calculated energies of the pygmy states are shown in the middle
panel. The values of the ratio m$_{1,{\rm LOW}}$/m$_{1,{\rm
HIGH}}$, of the energy weighted moments m$_1$ in the low-energy
region (E$\leq$10 MeV) and in the region of giant resonances
(E$>$10 MeV), are plotted in the lower panel.} \label{fig10}
\end{figure}

\begin{figure}
\caption{
(a) The RQRPA isovector
dipole response in $^{22}$O calculated for different values of 
the parameter $\epsilon_d$ (\ref{epsilond}).
(b) Neutron number operator response in
$^{22}$O computed for four values of the cut-off energy parameter $E_{Cp}$.
(c) The position
of the spurious $1^-$ state in $^{22}$O and $^{120}$Sn as a
function of the $2qp$ cut-off energy parameter $E_{Cp}$. 
(d) The excitation energies of the ISGMR in $^{22}$O and $^{120}$Sn
as functions of the cut-off energy parameter $E_{Ca}$. See text for 
description.} 
\label{fig11}
\end{figure}

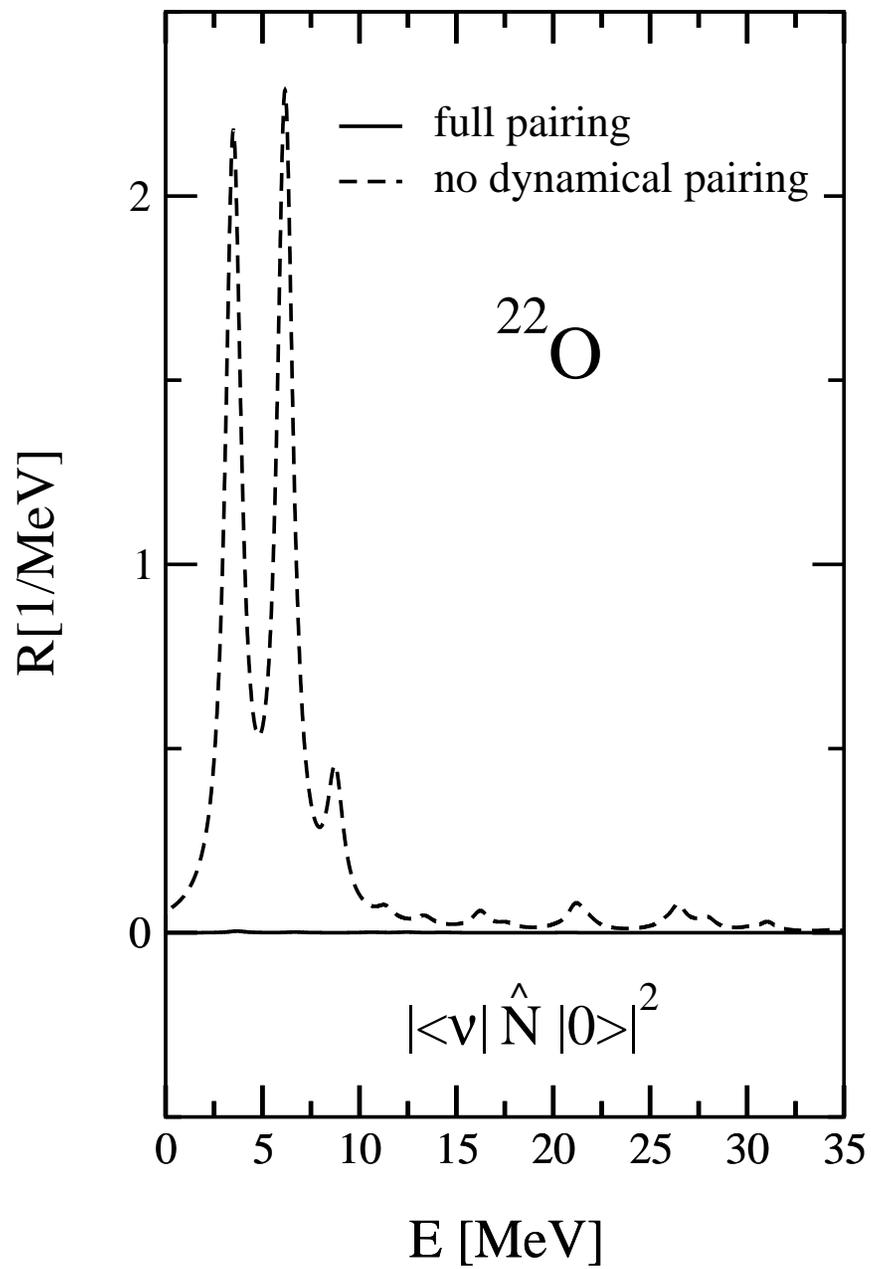
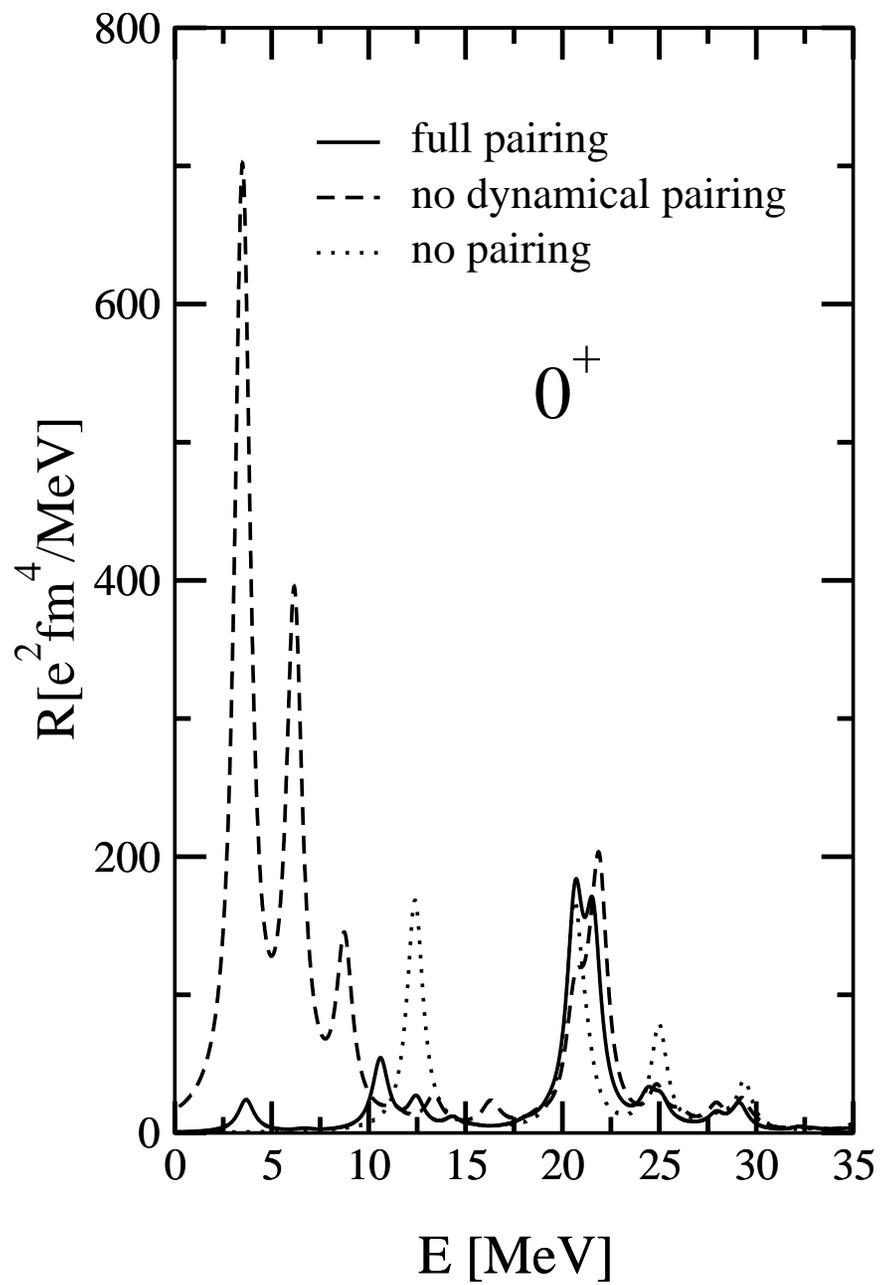
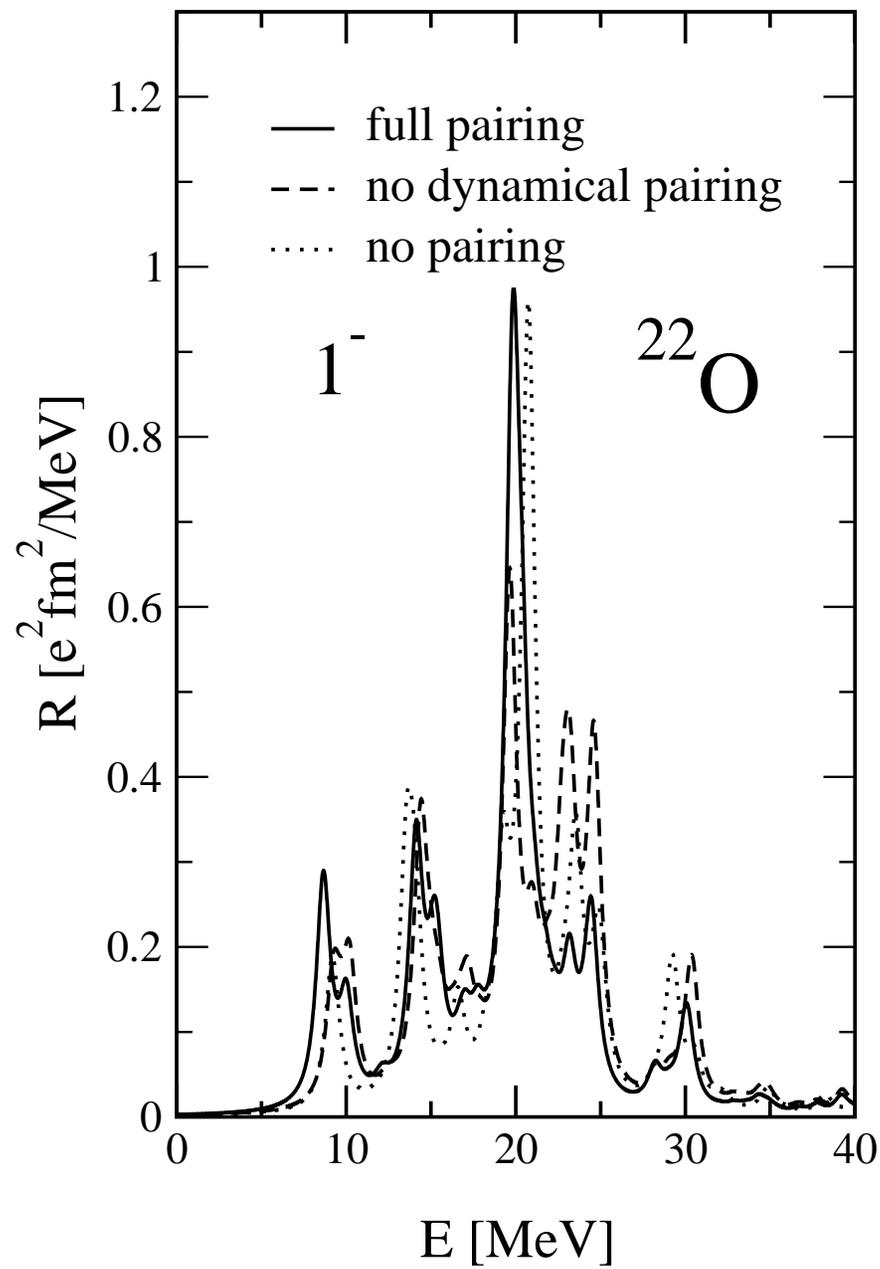
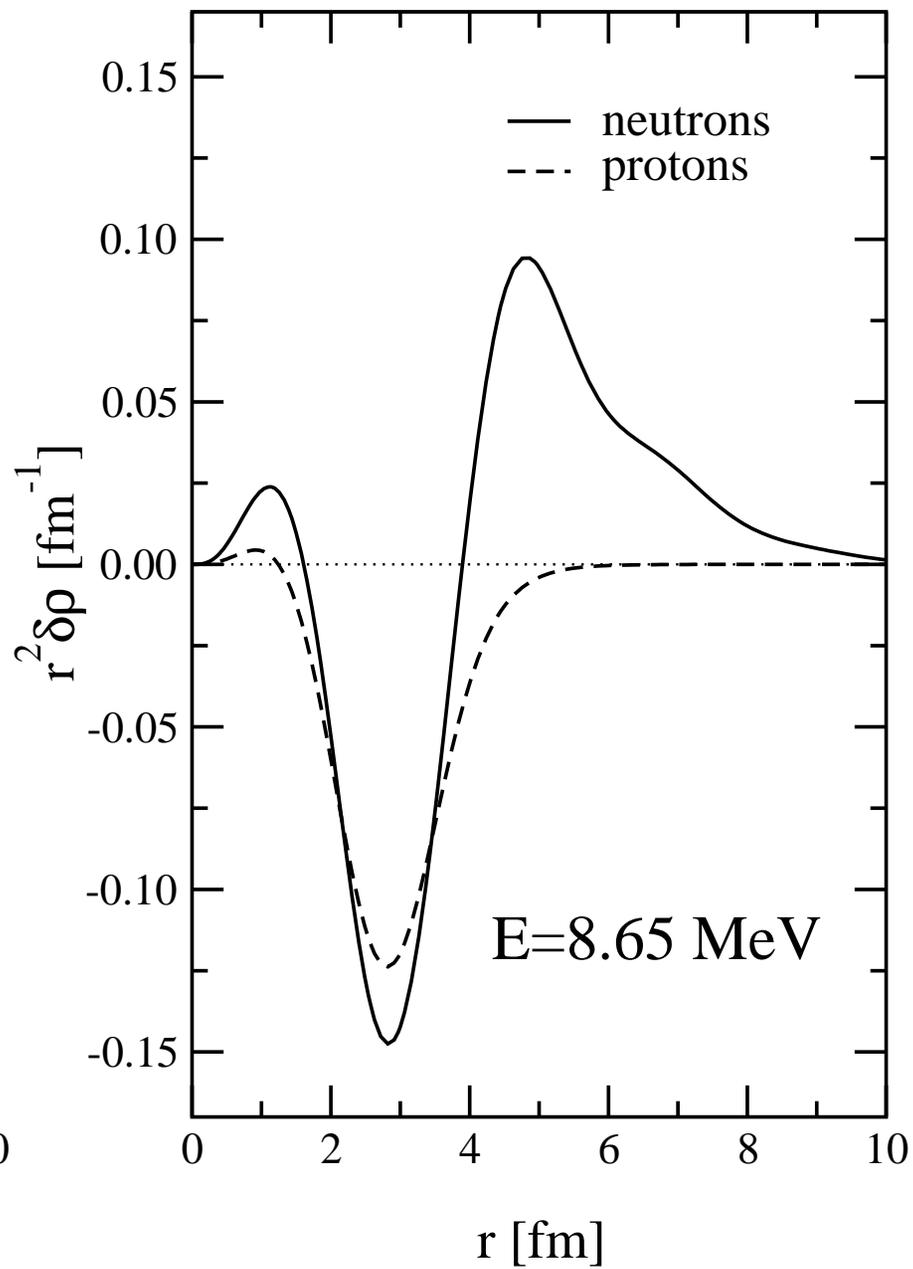
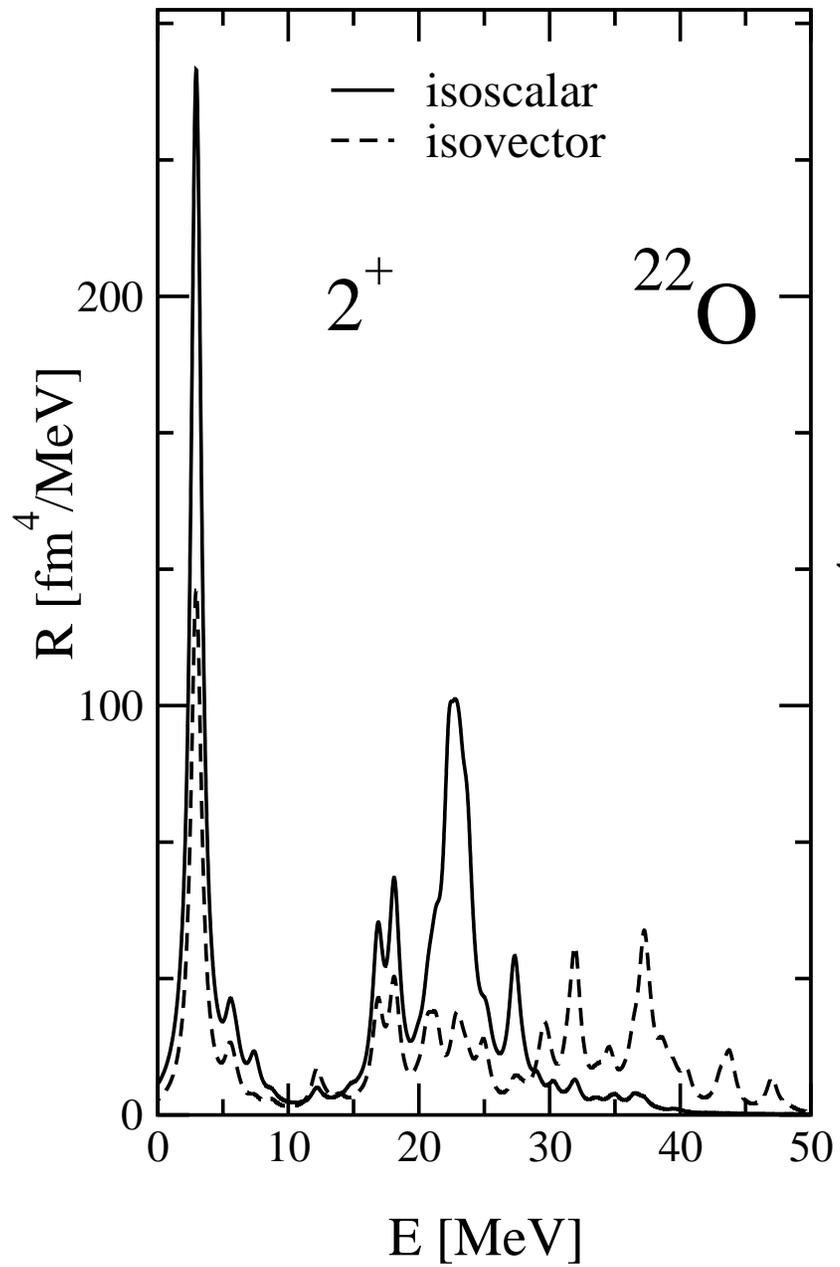
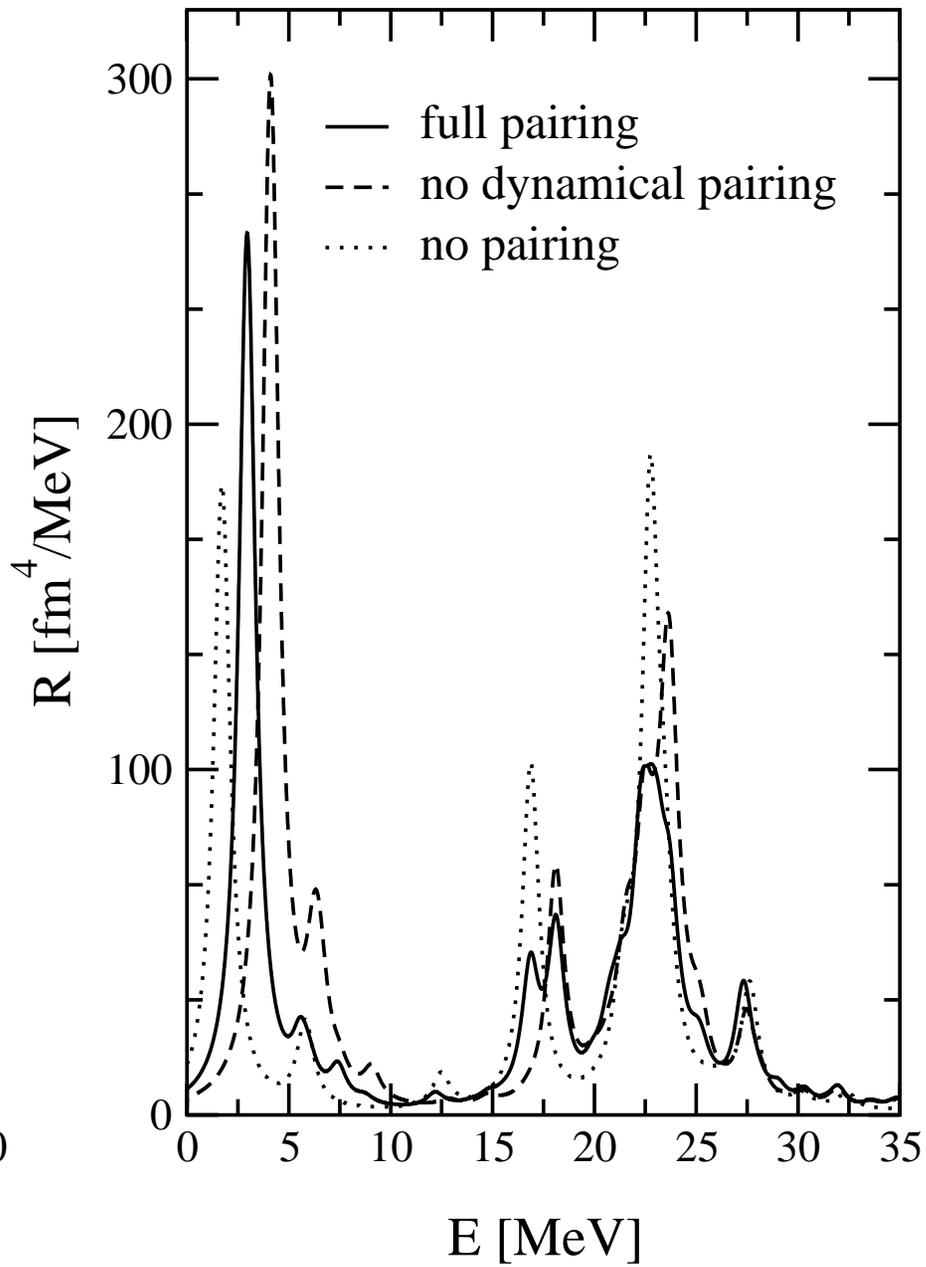
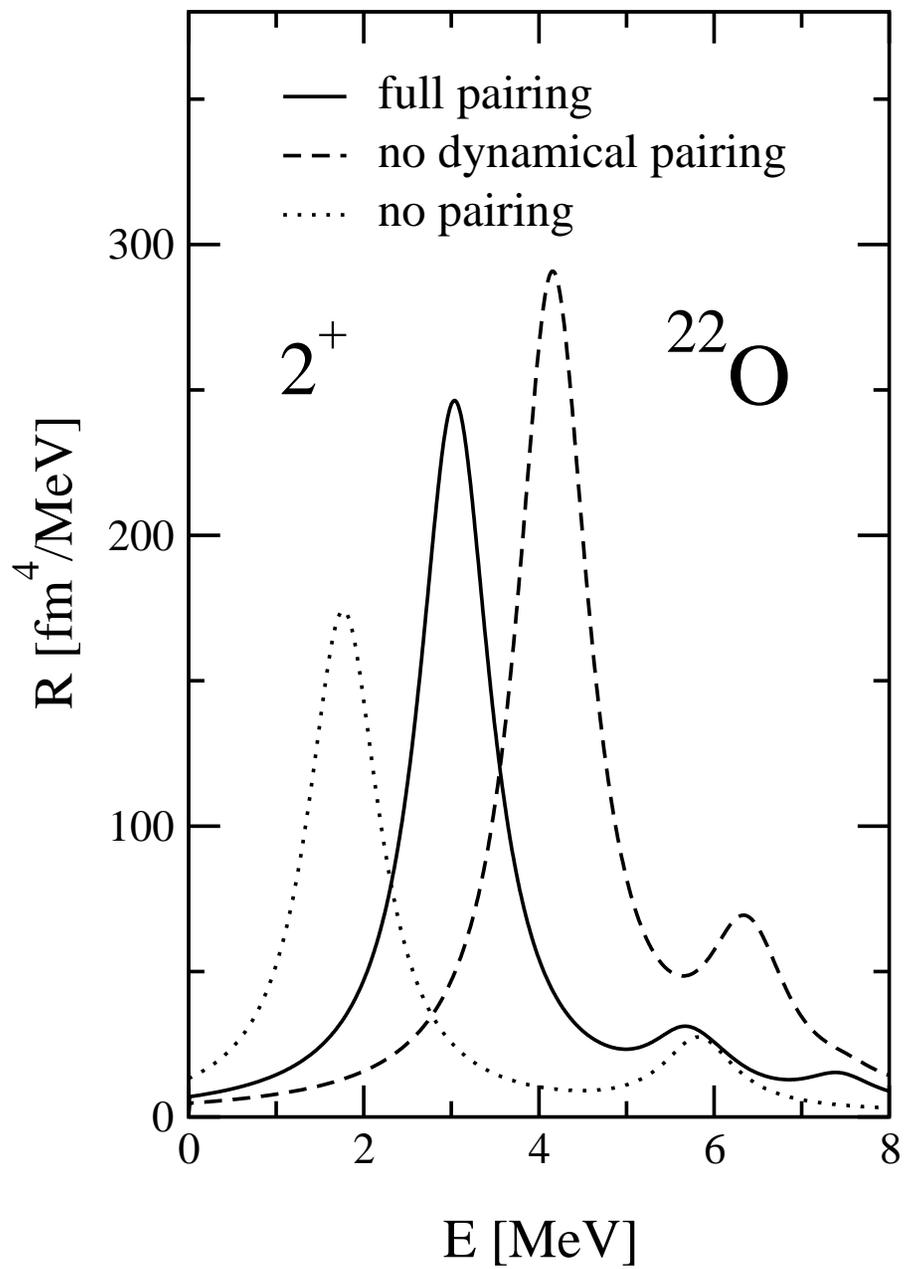
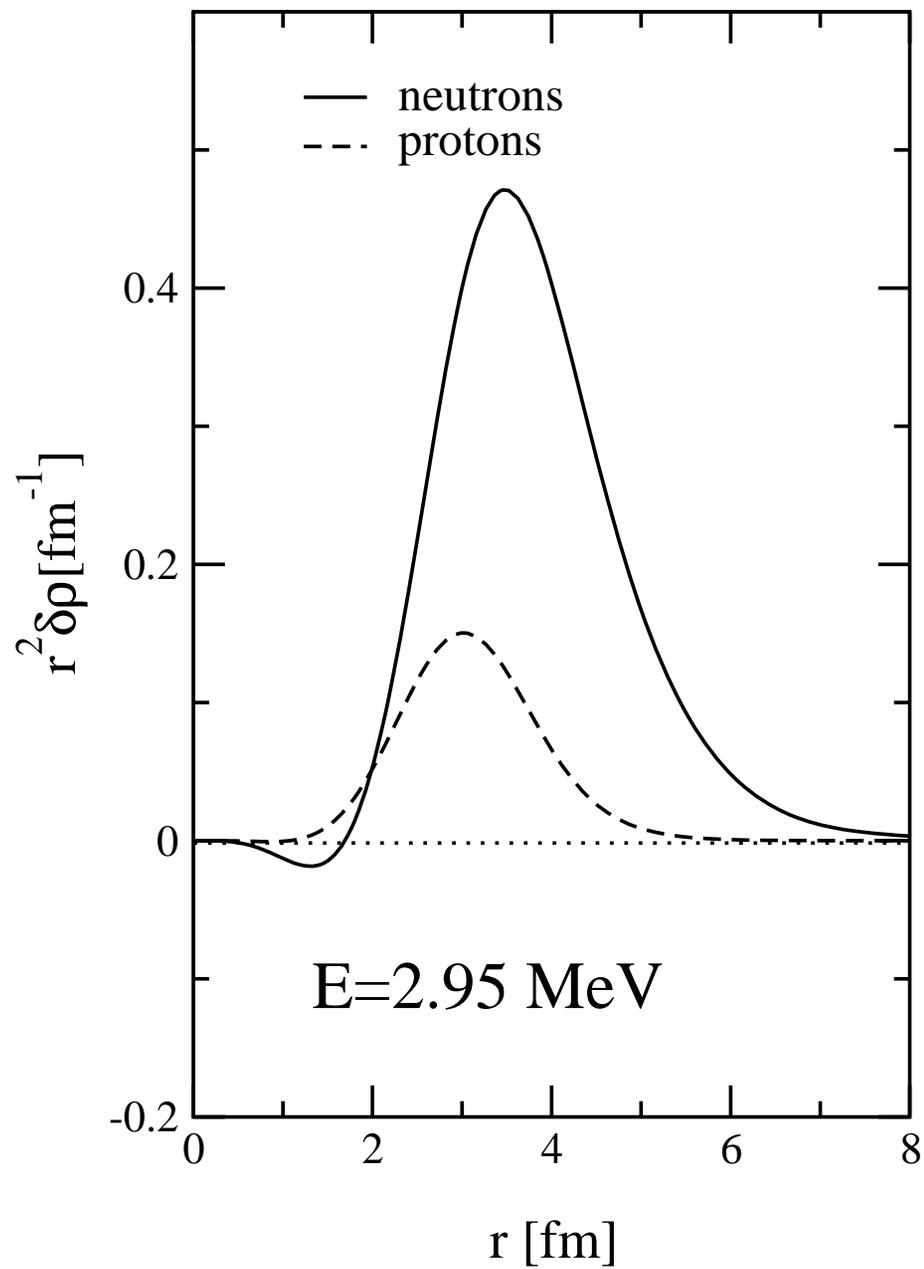
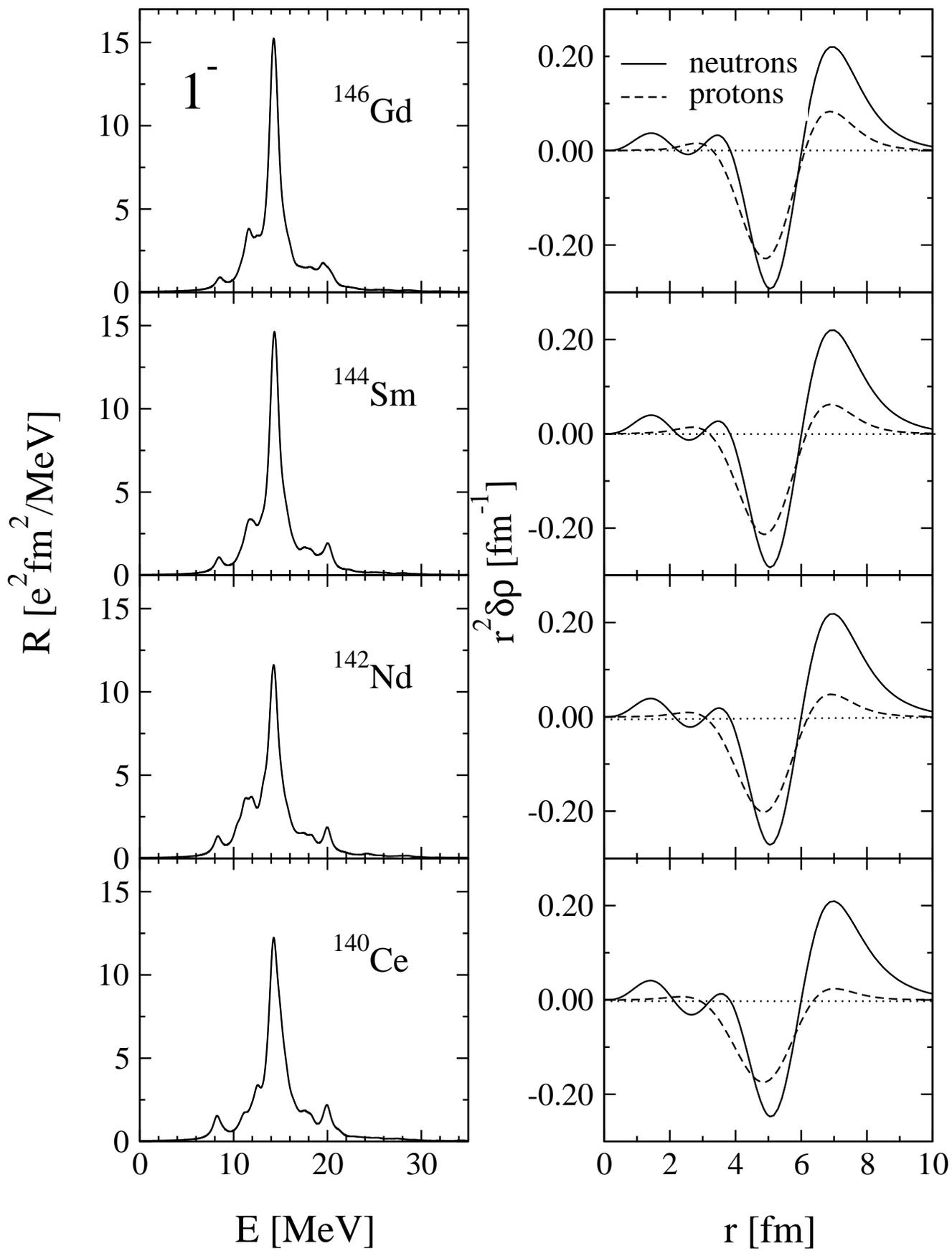
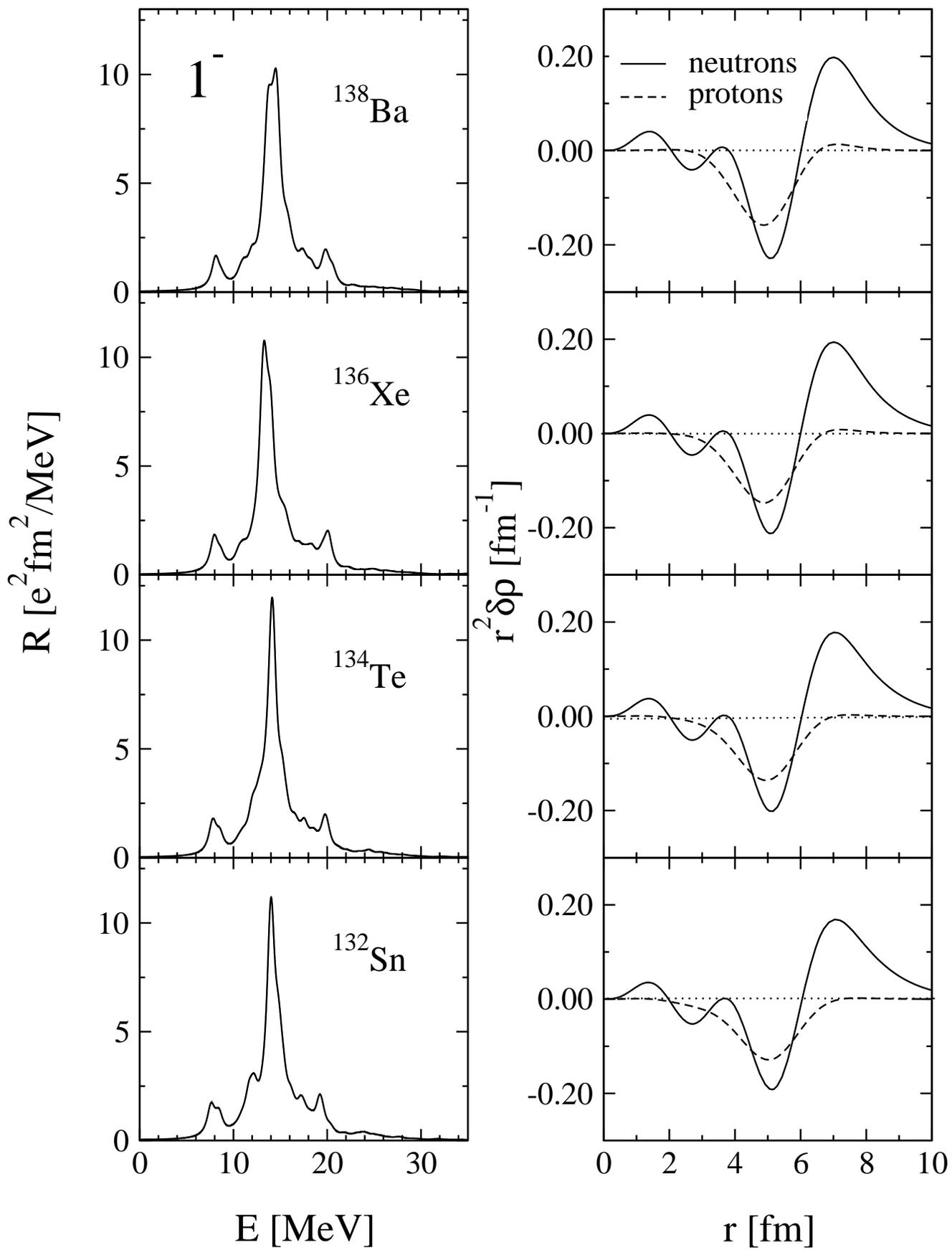
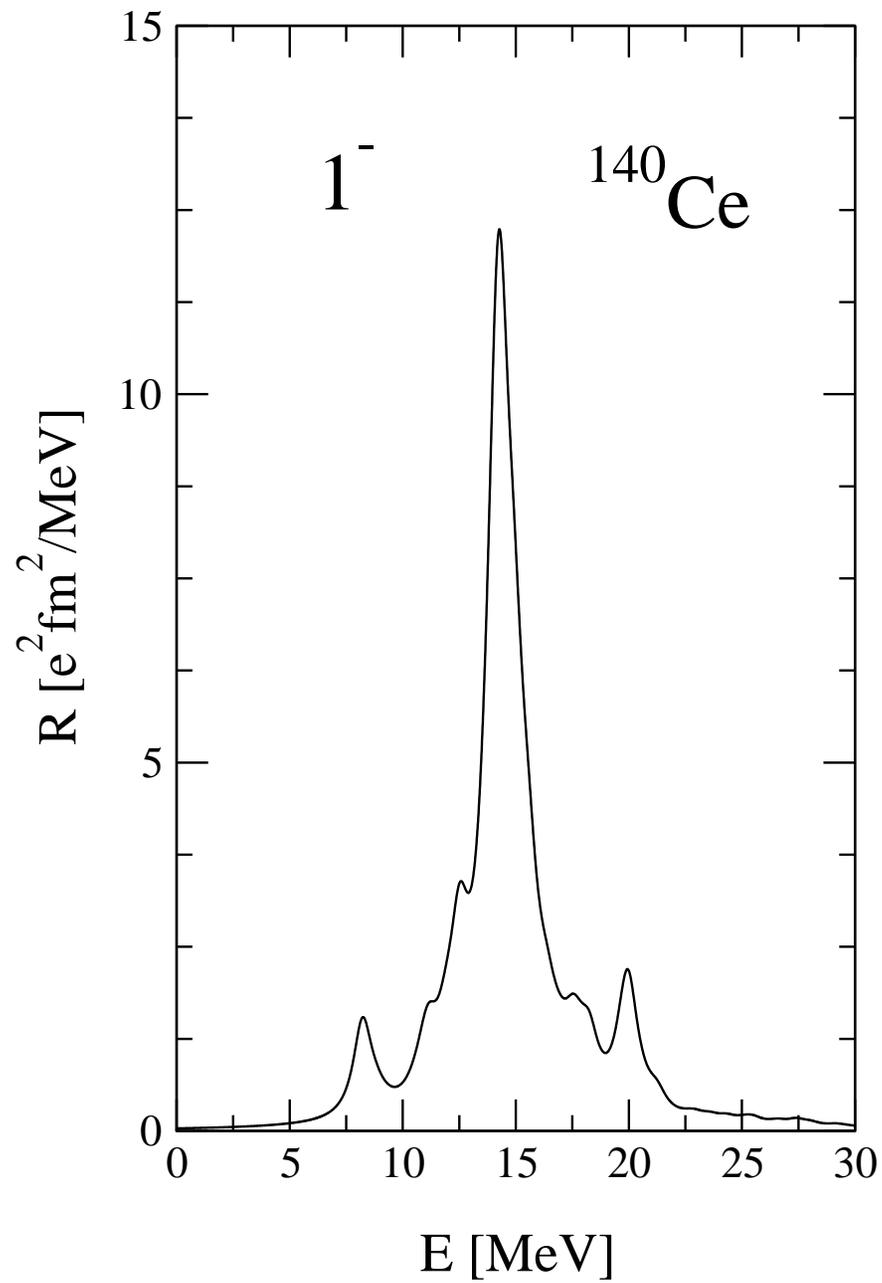
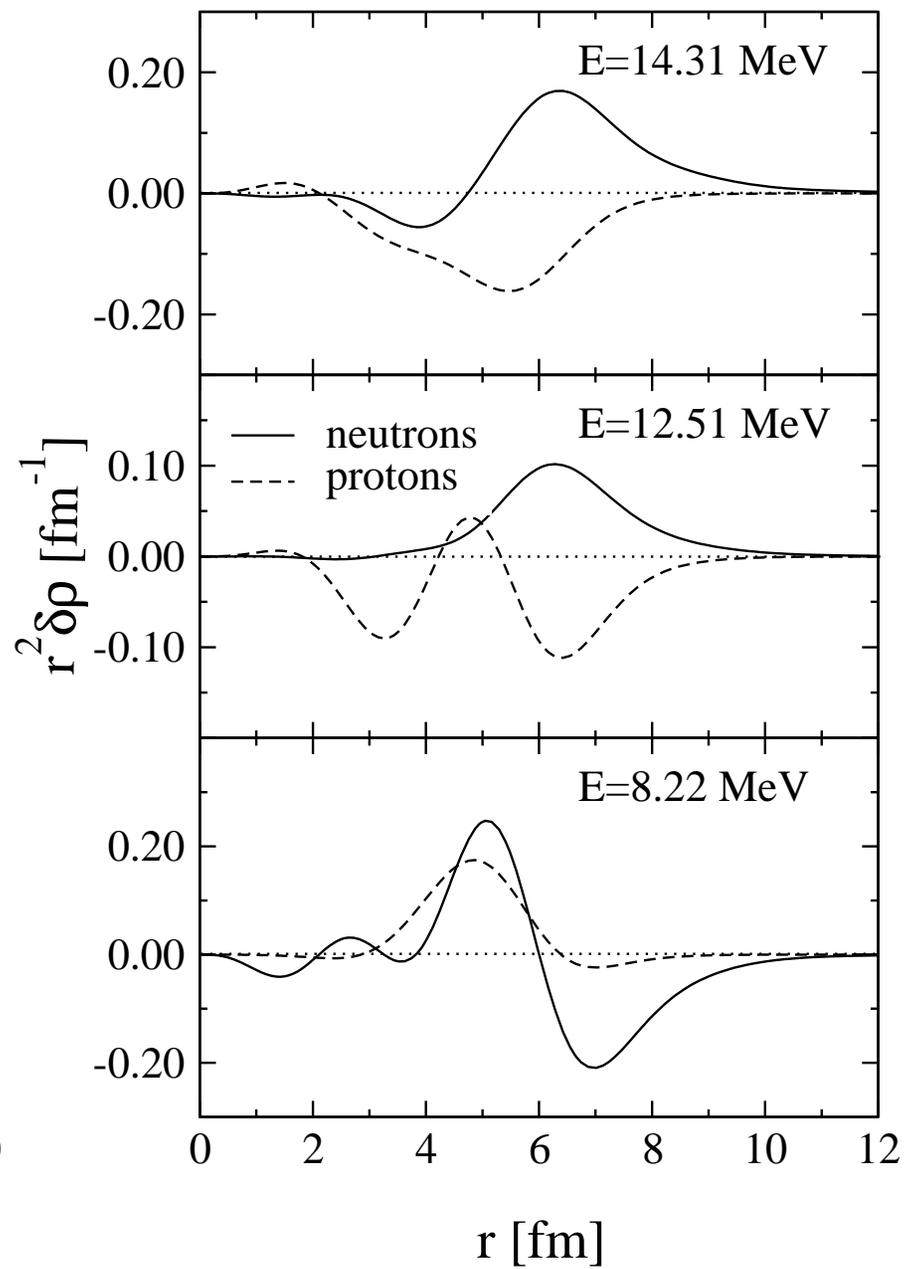
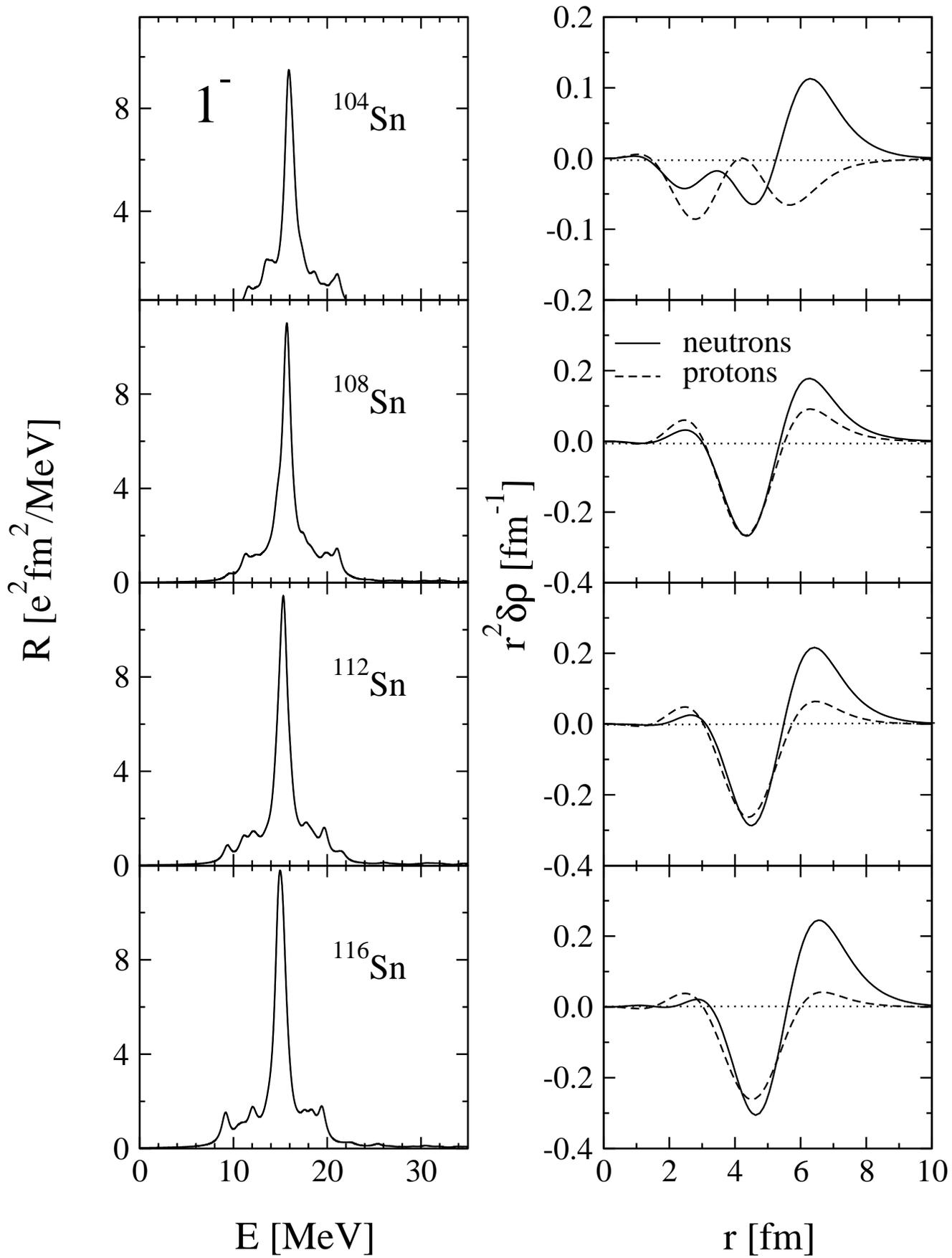
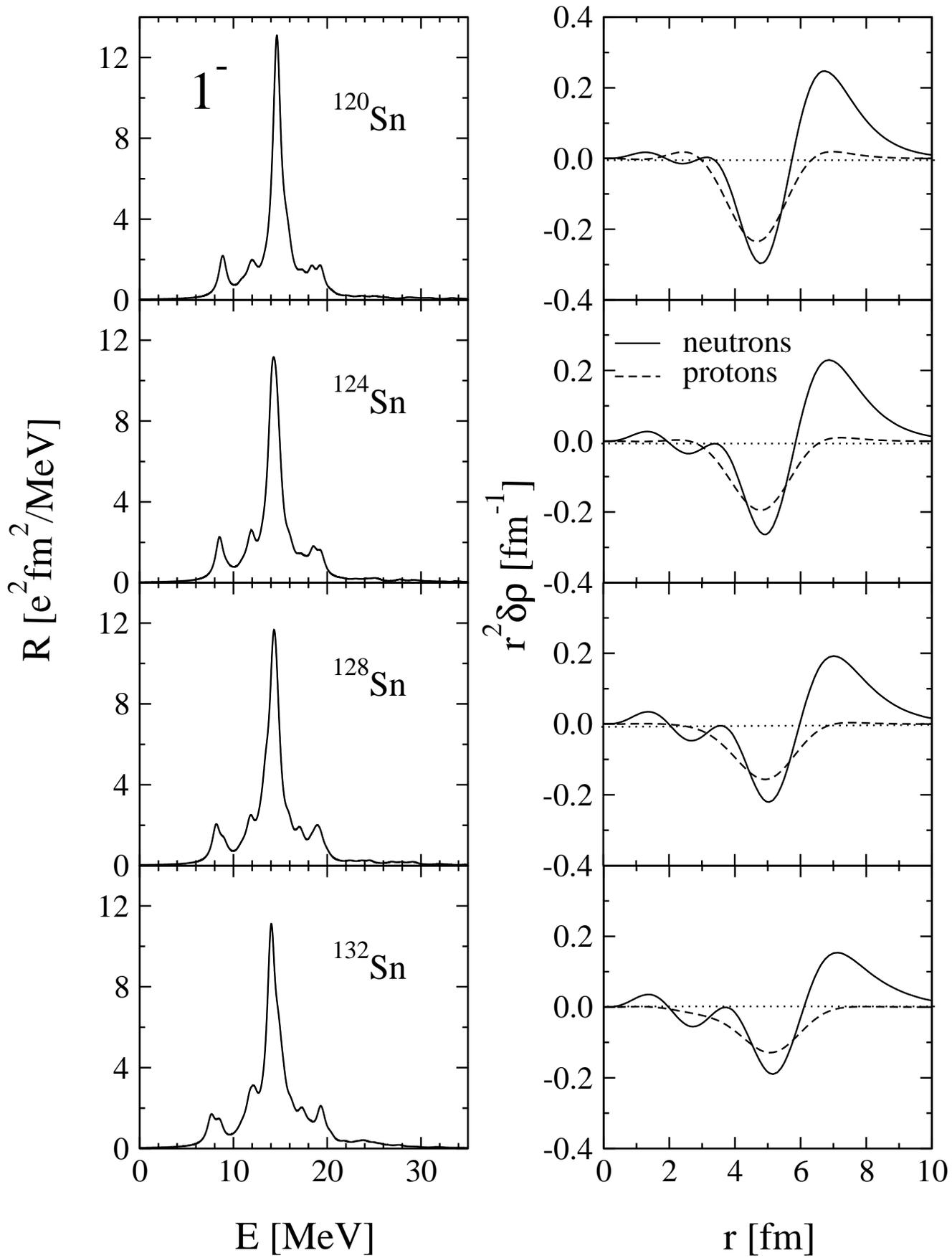
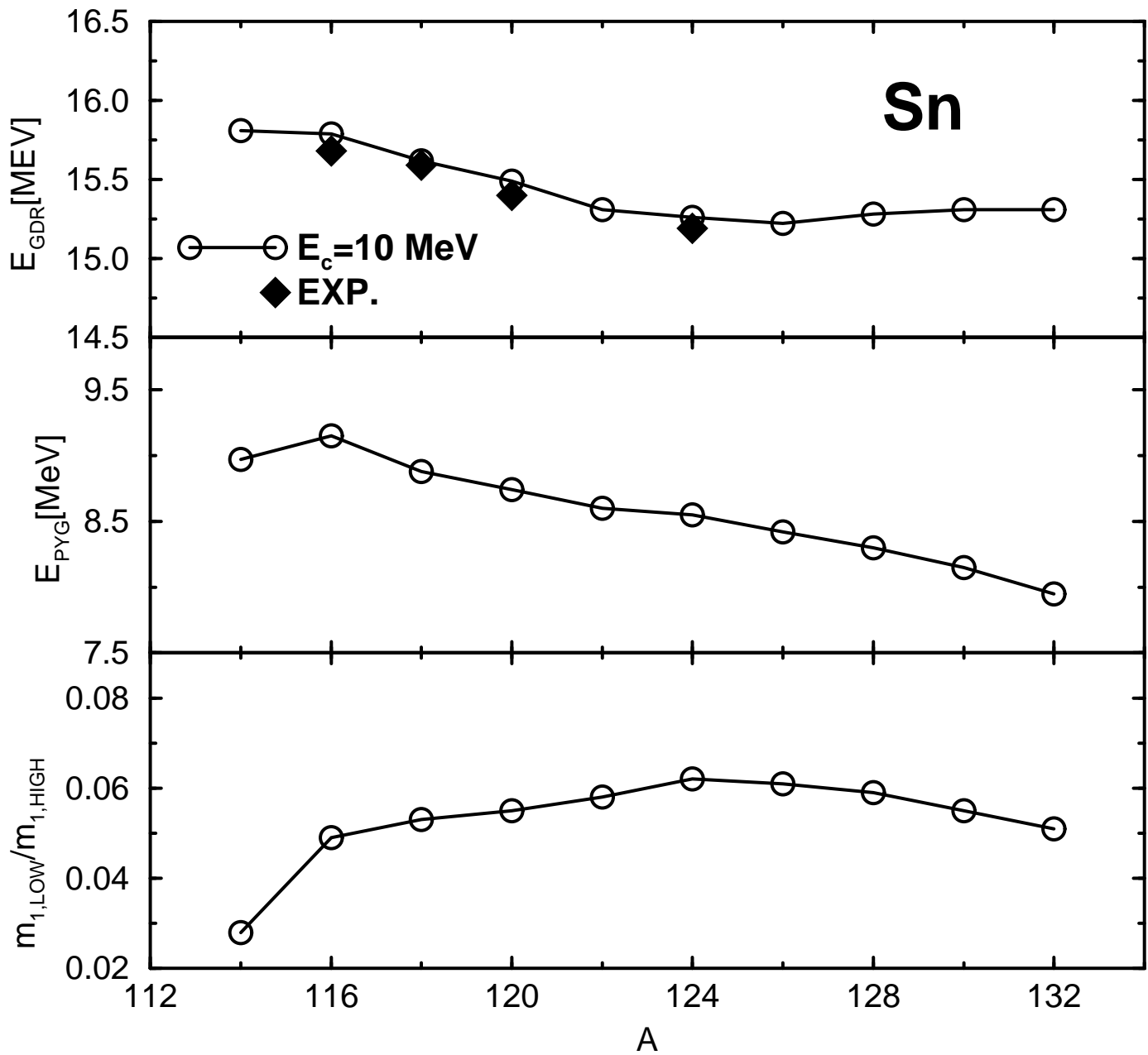
\end{document}